\documentclass[useAMS,usenatbib]{mn2e}
\usepackage{amsfonts,amssymb}
\usepackage{graphicx}
\usepackage{ulem}
\usepackage{lscape}
\begin{document}

\title[]{The Formation of Bimodal Dust Species in Nova Ejecta I: Chemical Conditions}
\author[ Zhu et al. ]{Chunhua Zhu$^{1}$\thanks{E-mail:chunhuazhu@sina.cn}, Helei Liu$^{1}$, Guoliang L\"{u}$^{1}$,
Zhaojun Wang$^{1}$, Lin Li$^{1}$\\
$^1$School of Physical Science and Technology, Xinjiang University, Urumqi, 830046, China\\}
%\altaffiltext{3}{Zentrum f\"{u}r Astronomie, Institut f\"{u}r
%Theoretische Astrophysik, Universit\"{a}t Heidelberg,
%Albert-\"{U}berle-Str. 2, D-69120 Heidelberg, Germany.}

\date{\today}

%\pagerange{\pageref{firstpage}--\pageref{lastpage}} \pubyear{2007}

%\maketitle

%\label{firstpage}

\pagerange{\pageref{firstpage}--\pageref{lastpage}} \pubyear{}

\maketitle

\label{firstpage}

\begin{abstract}
The formation of bimodal dust species (that is, both of silicate and amorphous carbon dust grains are
observed in a nova eruption) in nova ejecta is still debated.
Using the Modules for Experiments in Stellar Astrophysics code
and considering the effects of WD's mass, mass-accretion rate and the chemical profiles of WD which are described by new parameter --- mixing depth
on the chemical abundances of nova ejecta, we investigate the possibility that
bimodal dust species are produced in a nova eruption.
We find that
$C/O$ (the ratio of carbon number density to oxygen number density) of nova ejecta
is affected by the mixing depth. For the model with a small mixing depth,
the $C/O$ of nova ejecta can evolute from lager than 1.0 to smaller than 1.0 in a whole eruption, which
provides the chemical condition for the formation of bimodal dust species.

\end{abstract}
\maketitle
\section{Introduction}
It is well known, dust is the most important ingredient of
the interstellar medium. It is mainly produced by the stellar winds
of asymptotic giant branch (AGB) stars and supernova (SN) ejecta. Recently,
\cite{Lu2013} suggested that the dust produced by common-envelope (CE) ejecta
is not negligible \citep{Zhu2013,Zhu2015}.
\cite{Zhukovska2008} investigated the evolution of interstellar dust and stardust
in the solar neighbourhood, and found that the fraction of stardust produced
by SNe is about 15\% and it is about 85\% for AGB stars. \cite{Lu2013}
compared the dust masses produced by CE ejecta with those produced by AGB stars
for the solar metallicity, and found that the dust produced in CE ejecta may be
quite significant and could even dominate under certain circumstances. However, the former
greatly depends on the input parameters\citep{Lu2013}.

Based on the popular view of point,
due to the high binding energy of CO, the dust species produced by these sources
depend on the abundance ratio of the carbon to the oxygen ($C/O$) in their environment.
For example, M-type ($C/O<1$) AGB star can produce silicate dust grains
(olivine, pyroxene or quartz)\citep[e.g.,][]{Gail1999},
while amorphous carbon dust grains (graphite, diamond or silicon carbide) originate
 from C-type ($C/O>1$)  star\citep[e.g.,][]{Ferrarotti2005}. Of course, this expectation
 results from an assumption that the CO abundance reaches its saturation value.
 However, under some environment (such as SN, nova), there is a strong radiation
 field which can compromise CO molecule and reduce the criticality of the C/O.
 \citet{Pontefract2004} investigated the chemical
 evolution of nova ejecta, and found that amorphous carbon dust grains can be produced
 in an oxygen rich environment because of neutral reactions in a shielded region.

Similarly, dust has been observed in some nova eruptions \citep[e.g.,][]{Geisel1970,Gehrz1980}.
Although the dust produced by nova is less than 4\% of that produced by AGB stars\citep[e. g.,][]{Draine2009},
dust formation in the nova ejecta is very interesting.
About 50\% of novae can produce dust \citep{Harrison2018}.
Surprisingly, in some novae, both of silicate and  amorphous carbon dust grains
(that is, the bimodal dust species) are observed in a nova eruption
\citep{Gehrz1992,Evans1997}. Typically, the silicate grains are observed after amorphous carbon
grains are detected during nova eruptions. Due to no infrared echoes from the pre-existing silicate dust observed
in nova V1280 Sco, \cite{Sakon2016} considered that the silicate grains were newly produced during nova eruptions.
The bimodal dust species should be produced in a nova ejecta.
There is still debating about the origin of the bimodal dust species.
\cite{Sakon2016} suggested that the amorphous carbon grains form in the nova ejecta, but they
were not sure that the silicate grains were produced either in the expanding nova ejecta or
in the interaction zone of the nova ejecta and the oxygen-rich circumstellar medium.
\cite{Derdzinski2017} considered that CO is destroyed by the high energy particles accelerated by
the shock, and then the bimodal dust species can form.
For the former, V1280 Sco is a classical nova. This means that the companion of
white dwarf (WD) is a main-sequence star or dwarf
and its matter is transferred via Roche lobe. The circumstellar medium
does not originate from the companion but from the pre-nova ejecta.
Its composition should be similar with that of the nova ejecta.
For the latter, as \cite{Derdzinski2017} estimated,
the effects of non-thermal decomposition are very uncertain,
 so play a less  significant role.

Under the assumption of  saturated CO abundance,
the bimodal dust species mean that there should be two different chemical
environments which is noted by $C/O<1$ and $C/O>1$ in a nova ejecta.
The chemical abundances in the nova ejecta become the key to understand
the formation of the bimodal dust species.
However, it is well known that they strongly depend on the model of nova eruption.

The nova eruption is a thermonuclear runaway (TNR) occurring on the surface
of accreting WD in a close binary. It has been about four decades since
\cite{Starrfield1972} first used a nuclear reaction network to calculate
the TNR of a nova. The nova models have been investigated by many literatures
\citep[e.g.,][]{Prialnik1995,Jose1998,Yaron2005,Lu2008,Lu2009,Lu2011,Denissenkov2013,Denissenkov2014}.
See \cite{Starrfield2008}, \cite{Jose2008}, \cite{Starrfield2016} and \cite{Jose2016} for the recent reviews.
In these theoretical models, the chemical abundances in nova ejecta are mainly determined
by the TNR and the mixing from the WDs. Unfortunately, our knowledge of the mixing
is extremely limited. In general, in the 1D model, the range of the mixing is between 25\% and
75\%\citep[e. g.,][]{Jose1998,Starrfield2009}.
Recent multidimensional (2D and 3D) models
have showed that the Kelvin-Helmholtz hydrodynamic instabilities can
dredge up material from the underlying WD and
enrich the accreted envelope with (outer)core material\citep{Glasner2012,Casanova2016,Casanova2018}.
In addition,
most of models assume that WD has an uniform chemical composition.
However, we have known that WDs with different masses have different chemical compositions,
and the chemical abundances around the surface of a WD are deeply varietal with a depth increase.

In the present paper, we investigate the effects of the chemical profile
in WD and mixing depth on the chemical abundances in nova ejecta, and discuss
whether there is an environment for the formation of bimodal dust species in nova ejecta.
In \S  2, we give theoretical models about nova.
The main results are in \S 3. The conclusions are given in \S 4.

\section{Nova Models}
In nova models, the WD mass, the composition of the accreted matter,
the mass-accretion rate, convection, and the mixing prescription play an important role.
The WDs with different masses are produced
by stars with different masses, and they have undergone different
nucleosynthesis. Therefore, their chemical compositions are different.
We use \emph{Modules for Experiments in Stellar Evolution}
(MESA, [rev. 10108]; \cite{Paxton2011,Paxton2013,Paxton2015})
to create WD models which include the CO WDs with masses of 0.6 and 1.0 $\rm M_\odot$,
and ONe WDs with masses of 1.2 and 1.3 $\rm M_\odot$, respectively.
Usually, if WDs are produced via single star model,
there are some unburnt He- and H-rich layers above the CO cores or unburnt C-rich layers
above ONe cores \citep[e. g.,][]{Jose2016a}.  However, most of WDs in nova involve binary interaction including
Roche lobe mass transfer and CE evolution before they form \citep[e. g.,][]{Yungelson1993,Lu2006}.
\cite{Gil2003} showed that the binary interaction can greatly affect the WD masses and their chemical compositions.
These WDs probably are stripped H-rich layers, or even He-rich layers.
In \cite{Denissenkov2013}, the H-rich and He-rich layers of WDs removed artificially.
The Roche lobe mass transfer or CE evolution usually involve a donor
with H-rich envelope which is finally transferred to its companion or is ejected\citep{Eggleton2000,Nelson2001}.
Therefore, we remove H-rich layers artificially when WDs form in this work.

\begin{figure}

\includegraphics[totalheight=3.in,width=2.5in,angle=-90]{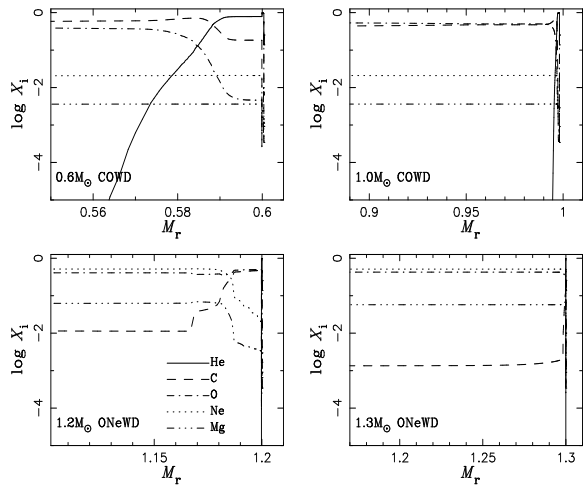}
\caption{The profiles of the chemical abundances for He, C, O, Ne and Mg around the
surface of WDs with different masses. Every WD is showed in  a sub-figure. The WD masses and
species are given in the bottom region. }
\label{fig:wdch}
\end{figure}

Figure \ref{fig:wdch} shows the profiles of the chemical abundances for He, C, O, Ne and Mg around the
surface of WDs with different masses. Obviously, the profiles of the chemical abundances for WDs with different masses are
different greatly. The main reason is that WDs with different masses originate from main-sequence stars with different masses
and they undergo different nuclear reactions.

In nova model, the mixing degree of the accreted matter with the matter of WD is
crucial parameter. In fact, during the progress of the TNR or the accretion, the mixing may occur.
There are several mechanisms for such mixing: \cite{Prialnik1984} assumed the mixing by
a diffusion layer\citep[e. g.,][]{Kovetz1985,Fujimoto1992};
\cite{Durisen1977} considered the mixing by shear instability due to differential rotation
\citep[e. g.,][]{MacDonald1983,Sparks1987,Livio1987};
\cite{Rosner2001} proposed the mixing by gravity waves\citep[e. g.,][]{Alexakis2002,Calder2002,Alexakis2004};
\cite{Woosley1986} suggested the mixing
by convective overshoot\citep[e. g.,][]{Shankar1992,Glasner1995,Glasner1997,Glasner2012}.
These mechanisms are put forward in the
framework of 1D or 2D simulations and have shortcomings of themselves\citep[][for details]{Livio1990}.
Considered that such convective mixing can only be simulated in the framework of three dimensions,
\cite{Casanova2011} carried  a 3D nuclear-hydrodynamic simulation for the mixing,
and found that Kelvin-Helmholtz instabilities can naturally result in
self-enrichment of the accreted envelopes with material from the underlying WD\citep[e. g.,][]{Casanova2016,Casanova2018}.
Their results are consistent with observations.
Following \cite{Denissenkov2013} and \cite{Rukeya2017}, we use 1D nova model in test suit of MESA to
simulate the nova eruption.
The main input parameters are listed as follows:\\
(I)\textbf{ The mixing depth}\\
Obviously, based on Figure \ref{fig:wdch}, the compositions of
TNR material will be different when the mixing occurs in different depth from the surface of WD.
Recently, in order to investigate the C-rich dust in CO nova outbursts, \cite{Jose2016a} also
considered the chemical profiles for the outer WD layers which are characterized by different C and O content .
In this work, we introduce a free parameter, mixing depth ($\delta=\frac{M_{\rm mix}}{M_{\rm WD}}$),
which is the ratio of the mixed mass of WD to the total mass.
In order to discuss the effect of the parameter $\delta$,
we take it as 0.001, 0.01, 0.05 and 0.1 in different simulations. \\
(II)\textbf{The nuclear network}\\ The element abundances of the accreted matter are similar with these of the Sun.
Because the temperature during the TNR can reach up to $2-4 \times 10^8$ K,
the nuclei as heavy as Ar and Ca may be synthesized.
In our model, we select 52 isotopes from $^1$H to $^{41}$Ca. These isotopes
refer to 386 nuclear reactions from pp chains, CNO cycle to Ca burning
(such as $^{41}$Ca($n,\alpha$)$^{38}$Ar, $^{41}$Ca($p,\alpha$)$^{38}$K), and so on.\\
(III)\textbf{The mass-loss rate}\\ The mass loss occurs when the luminosity of WD during the TNR closes to Eddington
luminosity. According to \cite{Denissenkov2013}, the mass-loss rate is given by
\begin{equation}
\dot{M}=-2\eta_{\rm Edd}\frac{L-L_{\rm Edd}}{v_{\rm esc}^2},
\end{equation}
where $v_{\rm esc}=\sqrt{2{\rm G}M/R}$ is the escape velocity,
$L$ and $L_{\rm Edd}=4\pi {\rm G c}M/\kappa$ are the luminosity of nova and Eddington luminosity, respectively.
Here, ${\rm G}$ and ${\rm c}$ are the gravitational constant and light velocity,
$M$ and $R$ are the WD's mass and radius, respectively. The $\kappa$ is the Rosseland mean opacity.
The parameter $\eta_{\rm Edd}$ is set to 1, which simply assumes that
the radiative energy of $L-L_{\rm Edd}$ is completely used to eject matter around the surface of WD.\\
(IV)\textbf{The mass-accretion rate and the core temperature}\\ It is widely accepted that the nova eruption is affected by
not only the WD's mass and the chemical abundances, but also the
mass-accretion rate and the core temperature of the WD. Here, we take different mass-accretion rates
($1\times10^{-7}$, $1\times10^{-9}$ and $1\times10^{-11} {\rm M}_\odot$ yr$^{-1}$) in different
simulations.
The effects of  the core temperature of WD on nova have been discussed
by many literatures \citep[e. g.,][]{Starrfield1998,Yaron2005,Jose2007}. In general, a cooler WD
can produce stronger nova outburst. In this work, we do not consider its effect, and
take a constant core temperature of $10^7$ K.

\section{Results and Discussions}
We simulate 48 models for nova eruption by combining 3 parameters (4 WD masses, 3 mass-accretion rates and
4 mixing depths). Tables 1---4 in appendix show all models and  results.

\subsection{Parameter Effects}
\begin{figure}
\begin{tabular}{c}
\includegraphics[totalheight=3.in,width=2.in,angle=-90]{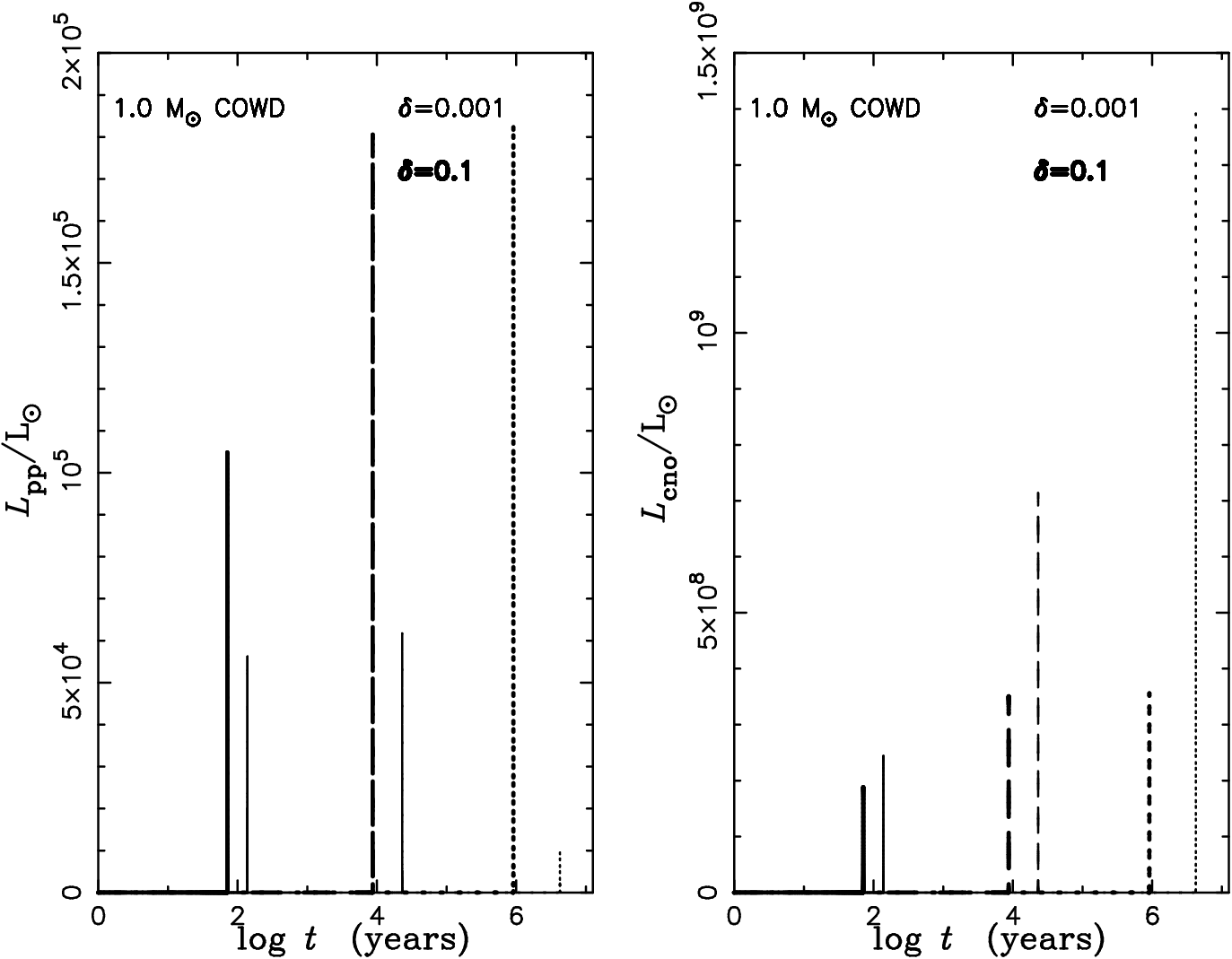}\\
\includegraphics[totalheight=3.in,width=2.in,angle=-90]{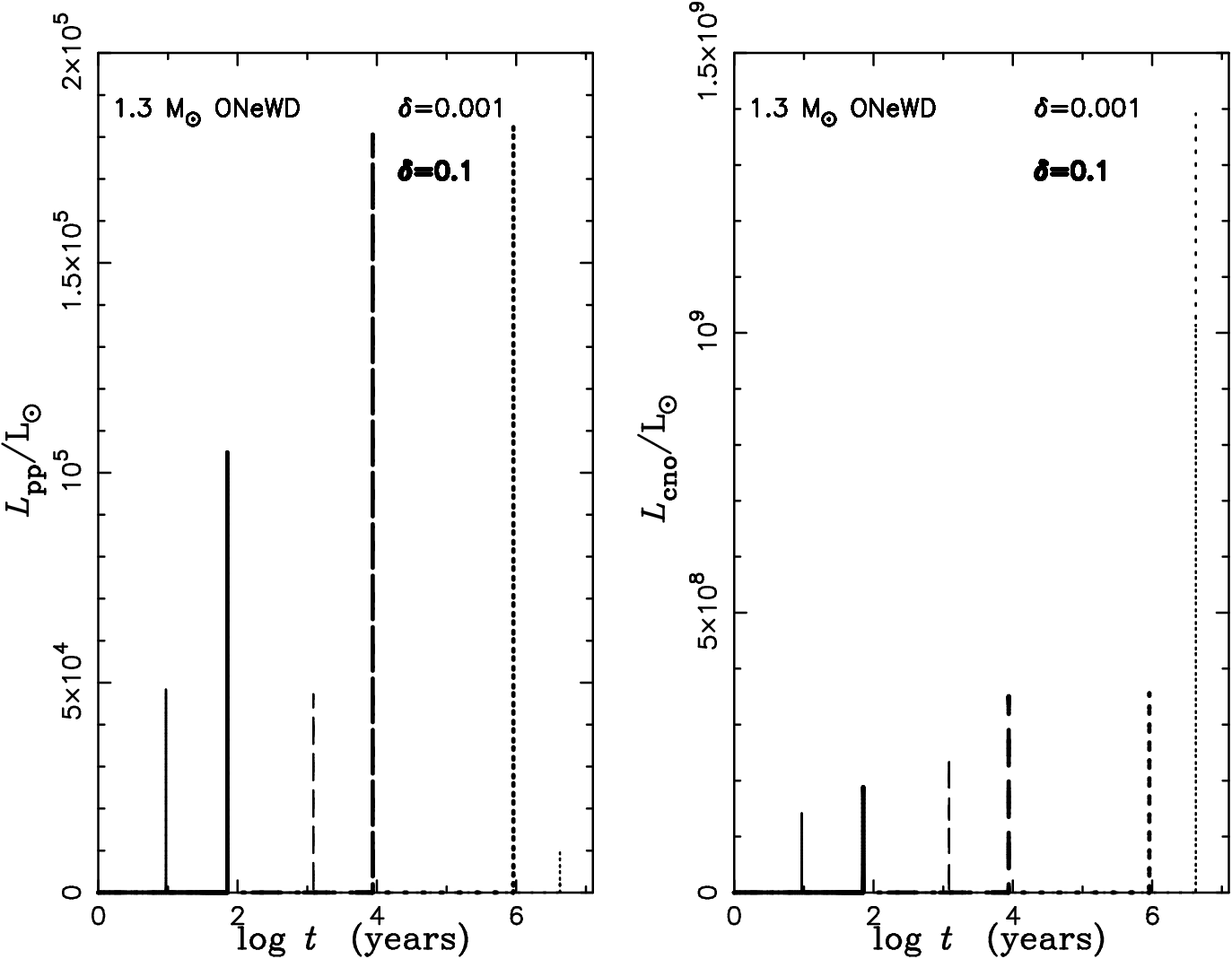}
\end{tabular}
\caption{The energy released by different nuclear reactions. The left and right
panels give these produced by the p-p chain and CNO cycle, respectively.
The up and down two  panels are for models with 1.0  $\rm M_\odot$ CO WDs and 1.3  $\rm M_\odot$ ONe WDs, respectively.
The thin and thick lines represent the results of the models with
$\delta=0.001$ and 0.1, respectively. The solid, dashed and dotted lines
show the models with the mass-accretion rates of $1\times10^{-7}$,
$1\times10^{-9}$ and $1\times10^{-11} \rm M_\odot$ yr$^{-1}$,
respectively.  }
\label{fig:nelu}
\end{figure}

In our models, the matter accreted is hydrogen-rich. Therefore, the energy released during
TNR mainly originates from hydrogen burning.
Due to very high temperature ($>10^8$K) in the
reaction zone, the CNO cycle is the main way for hydrogen burning, which is showed
in Figure \ref{fig:nelu} for typical models.
Simultaneously, WD masses, the mass-accretion rates and the mixing depth parameter ($\delta$)
have great effects on TNRs.
With similar previous studies \citep[e. g.,][]{Prialnik1995,Yaron2005}, the lower mass-accretion
rate of a WD is, the stronger TNR is, and a higher WD's mass is, the shorter and the stronger nova eruption is.
A large mixing depth can trigger an earlier TNR because it can provide more C and O elements for the
hydrogen burning in CNO cycle. Therefore, in these models, the hydrogen mass burned is less than that
with a small mixing depth and the maximum temperature during TNR is also lower (See Tables 1---4 in Appendix).

\begin{figure}
\begin{tabular}{c}
\includegraphics[totalheight=3.in,width=2.in,angle=-90]{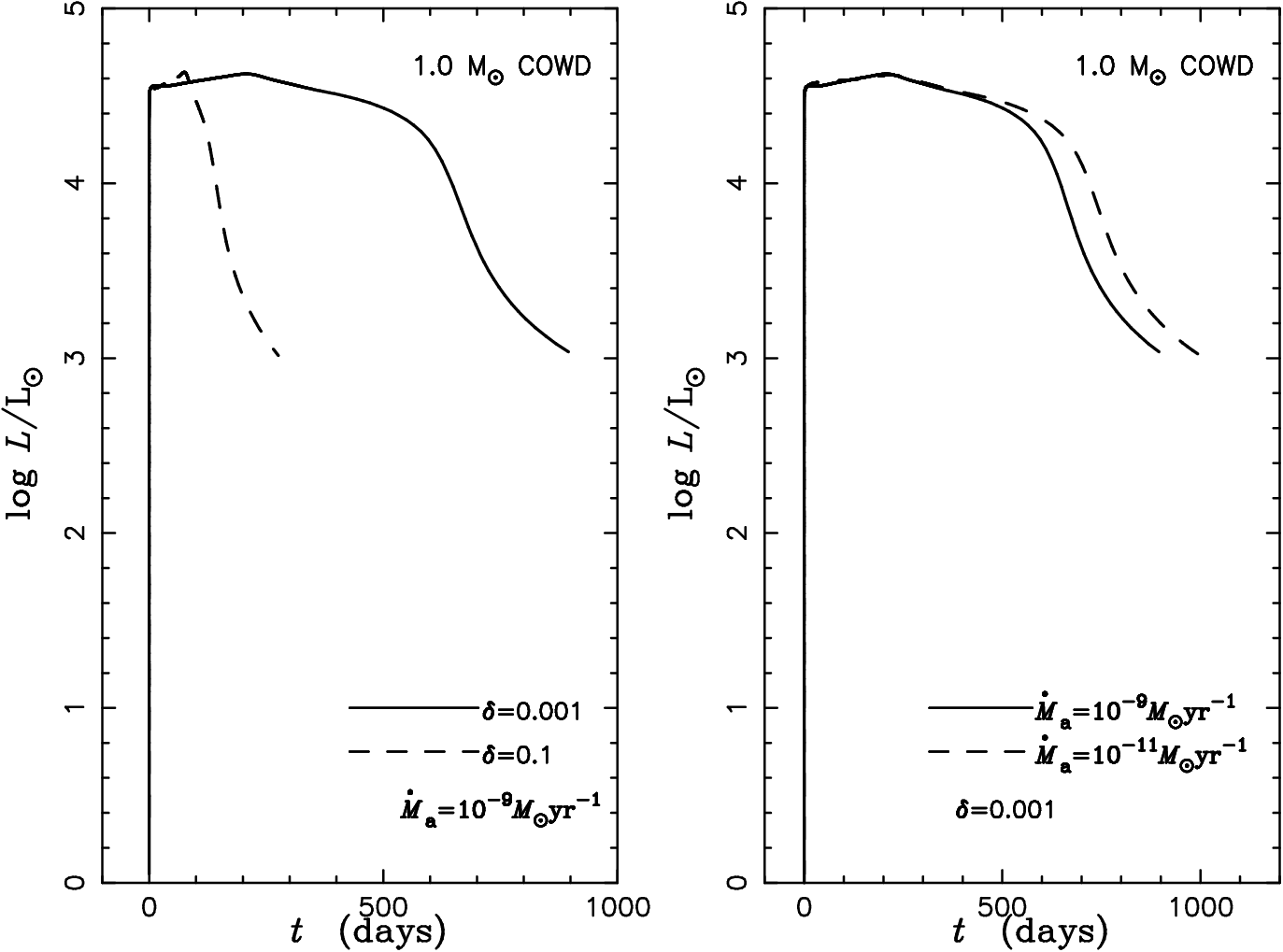}\\
\includegraphics[totalheight=3.in,width=2.in,angle=-90]{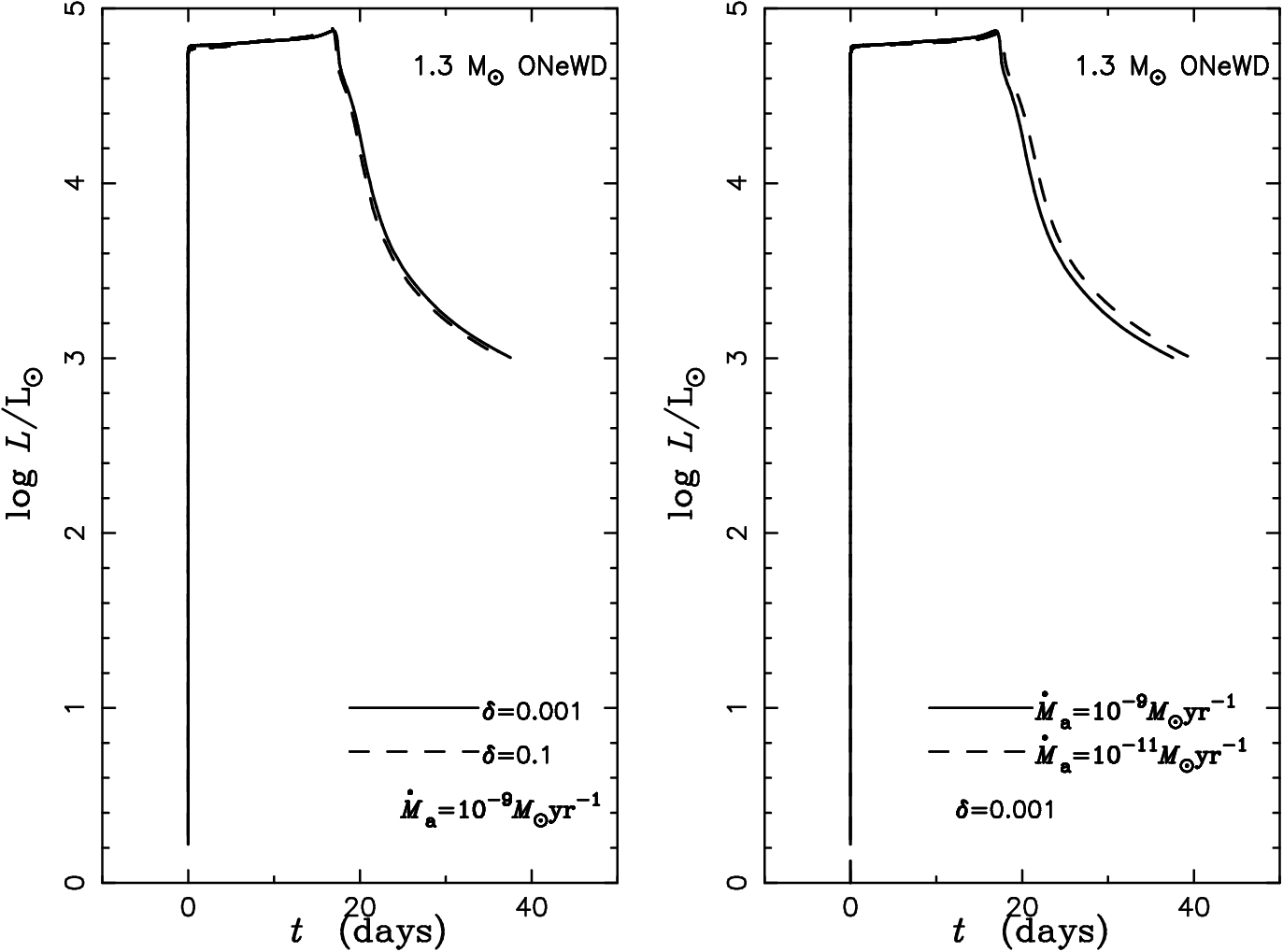}
\end{tabular}
\caption{The evolution of luminosity during a whole nova eruption in different models.
The up and down two  panels are for models with 1.0 $\rm M_\odot$ CO WDs and 1.3  $\rm M_\odot$ ONe WDs, respectively.
The left panels are for the model with different $\delta$ ($\delta=0.001$ and 0.1) but with
a fixed mass-accretion rate ($1\times10^{-9} \rm M_\odot$ yr$^{-1}$),
while the right panels are for the model with different mass-accretion rates
($1\times10^{-9}$ and $1\times10^{-11} \rm M_\odot$ yr$^{-1}$)
but a fixed $\delta$ ($\delta=0.001$).}
\label{fig:lumt}
\end{figure}

Figure \ref{fig:lumt} shows the evolution of luminosity during a whole nova eruption.
In our models, the duration of a nova eruption greatly depends on the mixing depth and WD mass,
while it is weakly affected by the mass-accretion rates. It increases from about 100 days
to about 700 days when $\delta$ increases from 0.001 to 0.1.
On the observations,
a nova eruption can last several weeks or many months, even serval years.
Therefore, we are not able to constraint the value of mixing depth parameter.
In short, besides of the core temperature, the theoretical simulation of nova eruptions in this work
greatly depends on the uncertain three parameters: the WD mass, the mass-accretion rate and the mixing depth.

%\begin{figure}
%\includegraphics[totalheight=3.5in,width=3.5in,angle=-90]{newnvev.ps}
%\caption{}
%\label{fig:wdch}
%\end{figure}

\subsection{Element Abundances in the Ejecta}
\begin{figure*}
\includegraphics[totalheight=6.in,width=4.in,angle=-90]{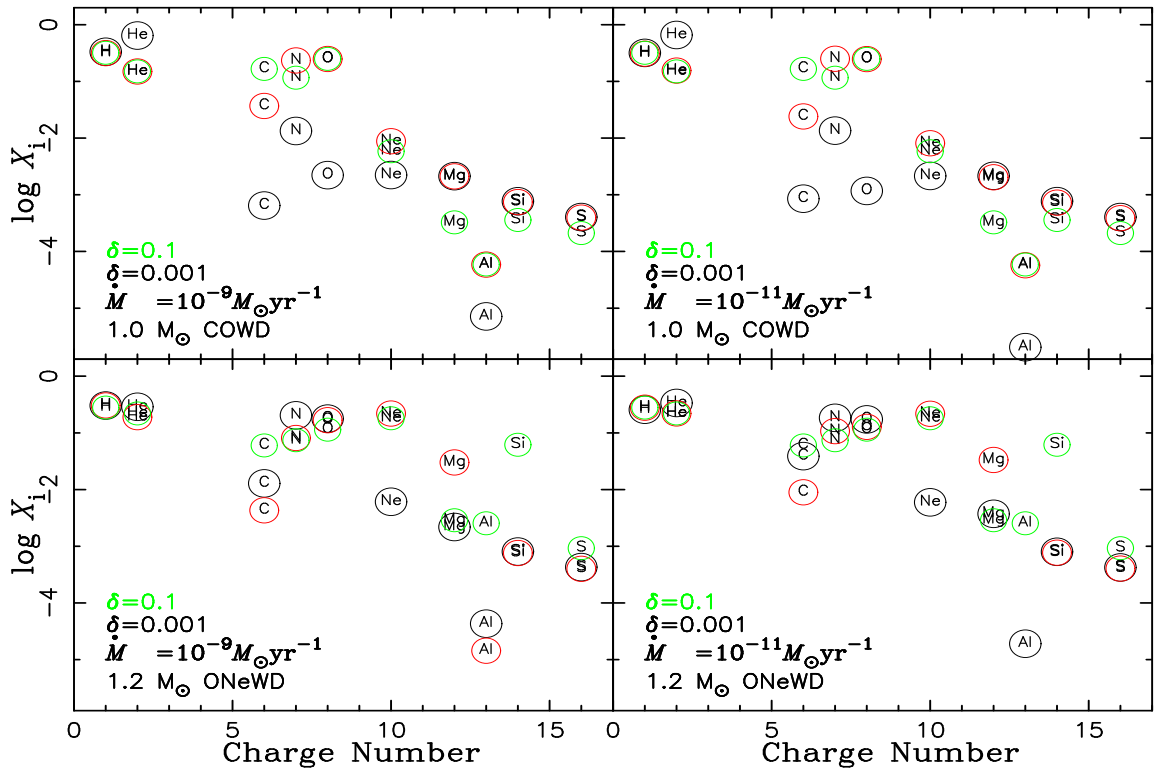}
\caption{The differences of chemical abundances for elements which can affect the formation of
dust in nova ejecta in models with different mass-accretion rates ($\dot{M}=10^{-9}$ and $10^{-11} \rm M_\odot$yr$^{-1}$)
and different mixing depths ($\delta=0.001$ and 0.1).
The black and green cycles represent the average chemical abundances
of the ejecta in the models with different mixing depths ($\delta=0.001$ and 0.1) during a whole eruption, respectively.
The red cycles in up and down pannel represent the results of CO3 and ONe5 models from \citet{Jose1998}, respectively.
}
\label{fig:nvchi}
\end{figure*}

Our models assume that the WD accretes solar composition material.
 After the mixing of the accreted matter
with the matter of WD, some matter is ejected during nova eruptions.
Therefore, the chemical abundances of ejecta also are influenced by the mass-accretion rate,
the mixing depth and WD's type. Usually, TNR can trigger nucleosynthesis
up to the charge number $\sim $ 20\citep{Iliadis2001}. Considering
that H, He, C, N, O, Mg, Al, Si and S  may affect the formation of dust  in nova ejecta
and these elements can be compared with those in \cite{Jose1998},
we show the average chemical abundances of these elements in the ejecta in Figure \ref{fig:nvchi}.
Compared with the mass-accretion rate (Comparing the left two panels with the right two panels in Figure \ref{fig:nvchi}),
the mixing depth and the WD's type have greater effects on the chemical abundances of the ejecta.
\cite{Jose1998} assumed that CO WD is composed of 49.5\% of C, 49.5\% of O and 1\% of Ne,
and the initial composition of ONe WD comes from C burning nucleosynthesis calculations
from \cite{Ritossa1996}.
In our work, with the enhance of the mixing depth, more C, O or Ne are involved into TNRs.
Therefore, the results of the CO WD model with large mixing depth ($\delta$=0.1)
are closed to those of CO3 model in \cite{Jose1998} although some input parameters are different.
However, the results of ONe WD in two works are different because of the different chemical abundances of
ONe WD.

The variations of the element abundances from the
accreted matter to the mixed matter result from the mixing, and are determined by
the chemical profiles of WD and the mixing depth. The variations from
the mixed matter to the ejecta are triggered by the TNR.
Therefore, as Figure \ref{fig:nvchi} and Tables 1---4 show, for a low-mass CO WD, the mixing
can change the abundances of C, O and Mg while the TNR only varies the abundances of elements lighter
than O element because the maximum temperature during nova eruptions hardly gets to $\sim 2.0\times10^8$ K;
for a high-mass ONe WD, the elements lighter than Ca will be involved in some nucleosynthesises.
These results are consist with those of \cite{Jose1998}.

\begin{figure}
\includegraphics[totalheight=3.in,width=2.5in,angle=-90]{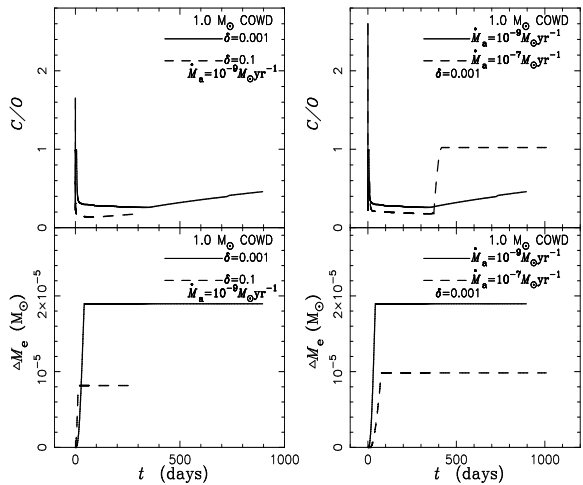}
\caption{Similar with Figure \ref{fig:lumt}, but for the evolutions of $C/O$ in nova ejecta and the
mass ejected ($\Delta M_{\rm e}$).}
\label{fig:cot}
\end{figure}

Due to very high binding energy of CO, the species of dust greatly depend on $C/O$
in the ejecta. Figure \ref{fig:cot} gives the $C/O$ evolutions in nova ejecta and the amount of mass ejected for
different models. Obviously, the $C/O$ of ejecta is larger than 1.0 at the beginning of the
nova eruption, while then quickly becomes smaller than 1.0 due to the nucleosynthesises of TNR.
As the bottom two panels of Figure \ref{fig:cot}  shows, about less than 10\% of mass ejected is carbon-rich.
It is possible that the amorphous carbon dust is firstly
formed, and soon the silicate dust is formed in the nova ejecta, which
is similar with the observations found by \cite{Gehrz1992,Evans1997,Sakon2016}.

\begin{figure}
\includegraphics[totalheight=3.5in,width=2.5in,angle=-90]{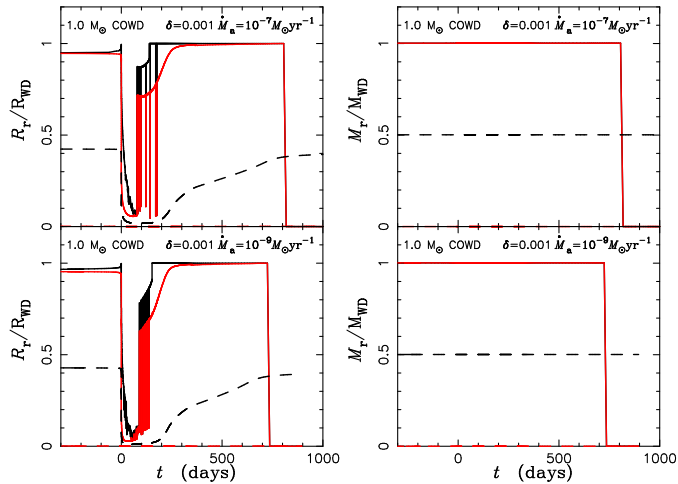}
\caption{The evolution of the convective and the mixing regions during nova eruptions in different models.
The beginning of the eruption is represented by $t=0$.
The left two panels show the convective and the mixing regions along the coordinate of WD's radius, while the right
two panels give those along the coordinate of WD's mass. The convective region lies between the black and red solid lines,
and the thermohaline mixing region is located between the black and red dashed lines. The input parameters of models
are given in the blanks.
}
\label{fig:conv}
\end{figure}

Figure \ref{fig:conv} shows the evolution of the convective and the mixing regions
during nova eruptions in different models. Before the outburst,
the convection produced by the accretion occurs in a very thin layer under the WD surface, and the mixing triggered by the
thermohaline mechanism always exists in the thick core region. During the outburst, the envelope accumulated on the
WD surface rapidly expands up to several hundred times of the WD radius, and the convection still occurs in the
bottom of the envelope. Based on the right two panels of Figure \ref{fig:conv}, compared with WD mass,
the mass involved in the convection region is insignificant, that is,
the mixing between the matter accreted and the WD's matter during the outburst process in our models
mainly occurs in a thin layer on the bottom of the envelope accumulated.

\begin{figure*}
\begin{tabular}{cc}
\includegraphics[totalheight=3in,width=2.in,angle=-90]{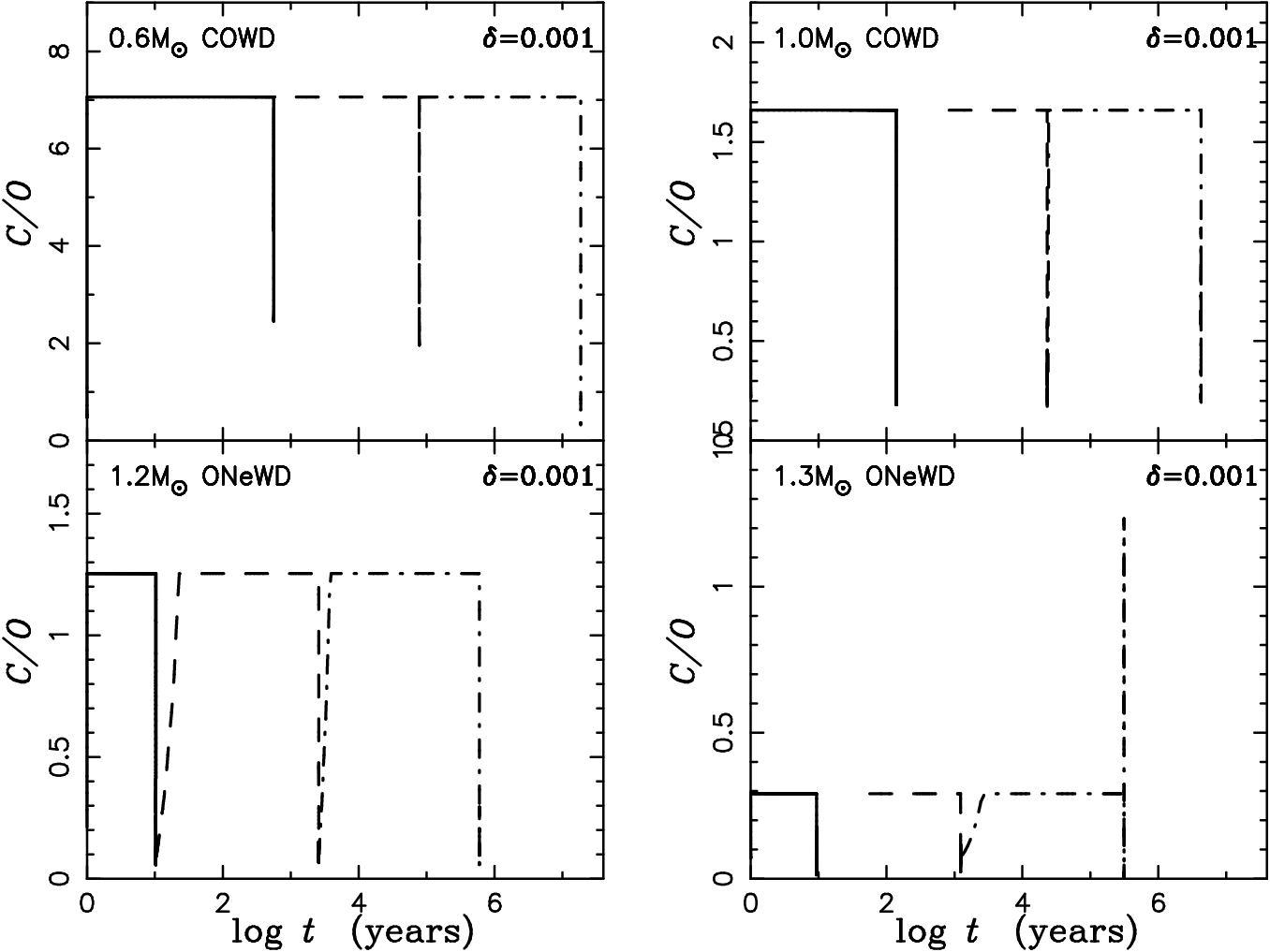}
&\includegraphics[totalheight=3in,width=2.in,angle=-90]{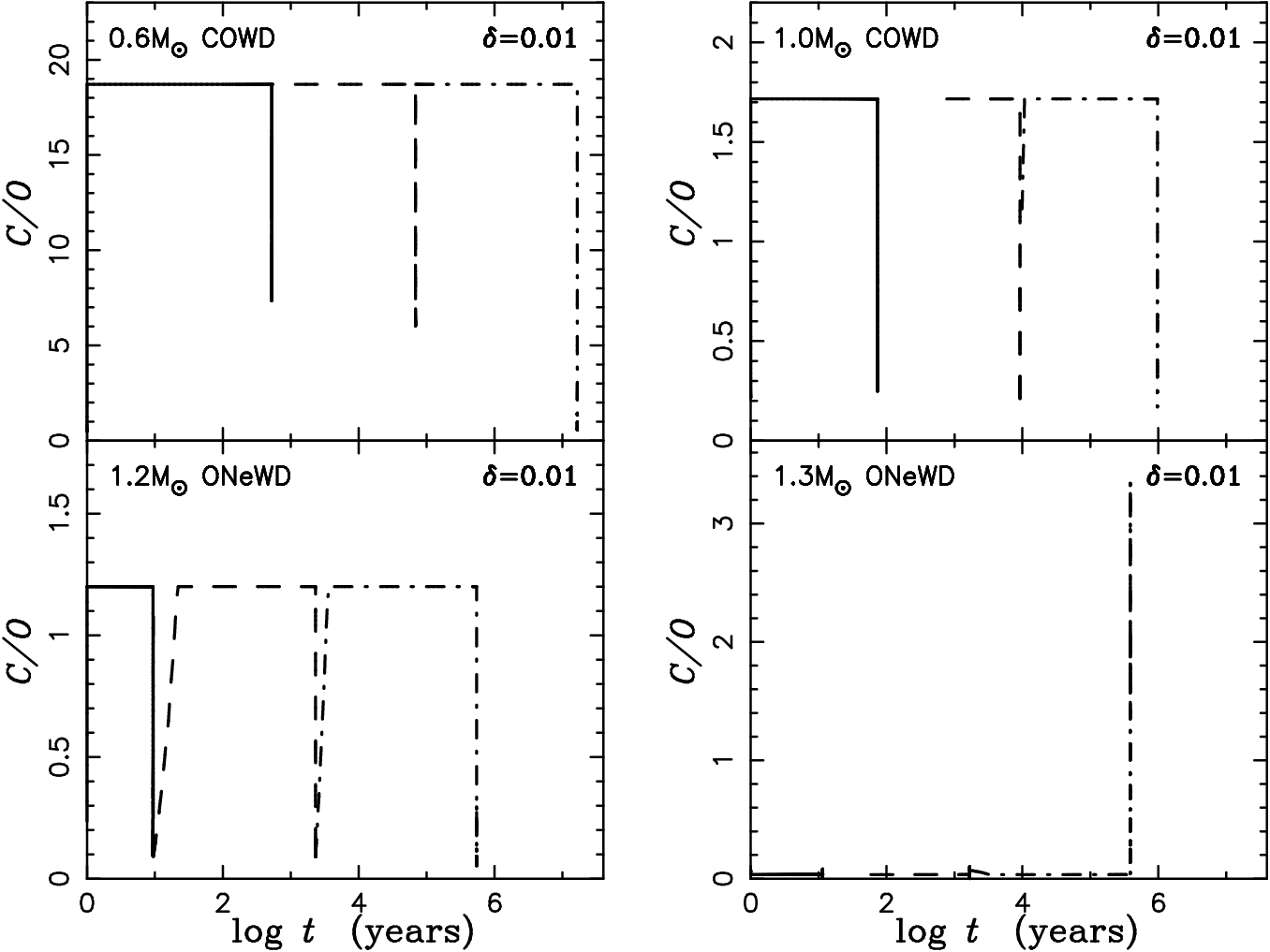}\\
\includegraphics[totalheight=3in,width=2.in,angle=-90]{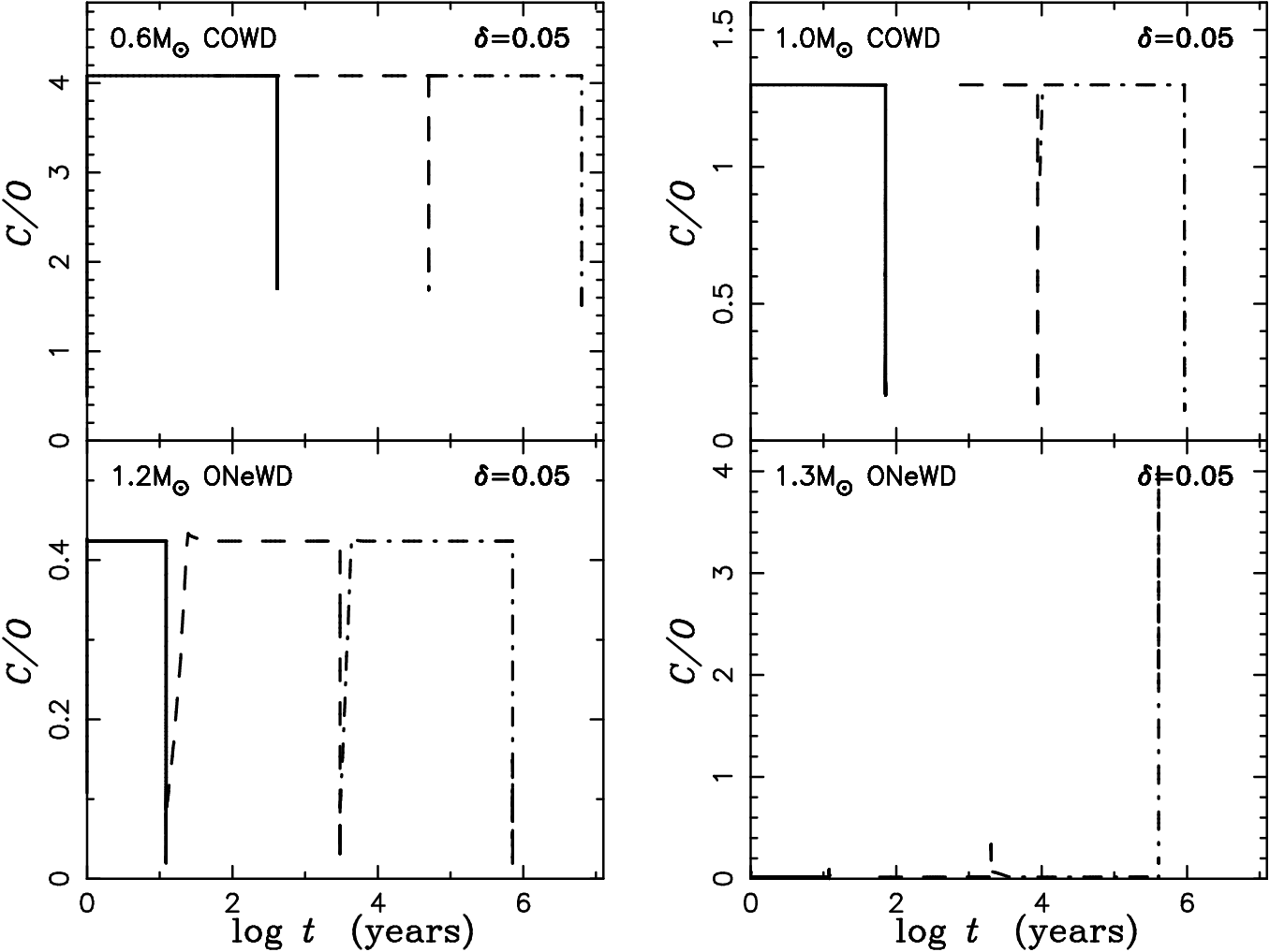}
&\includegraphics[totalheight=3in,width=2.in,angle=-90]{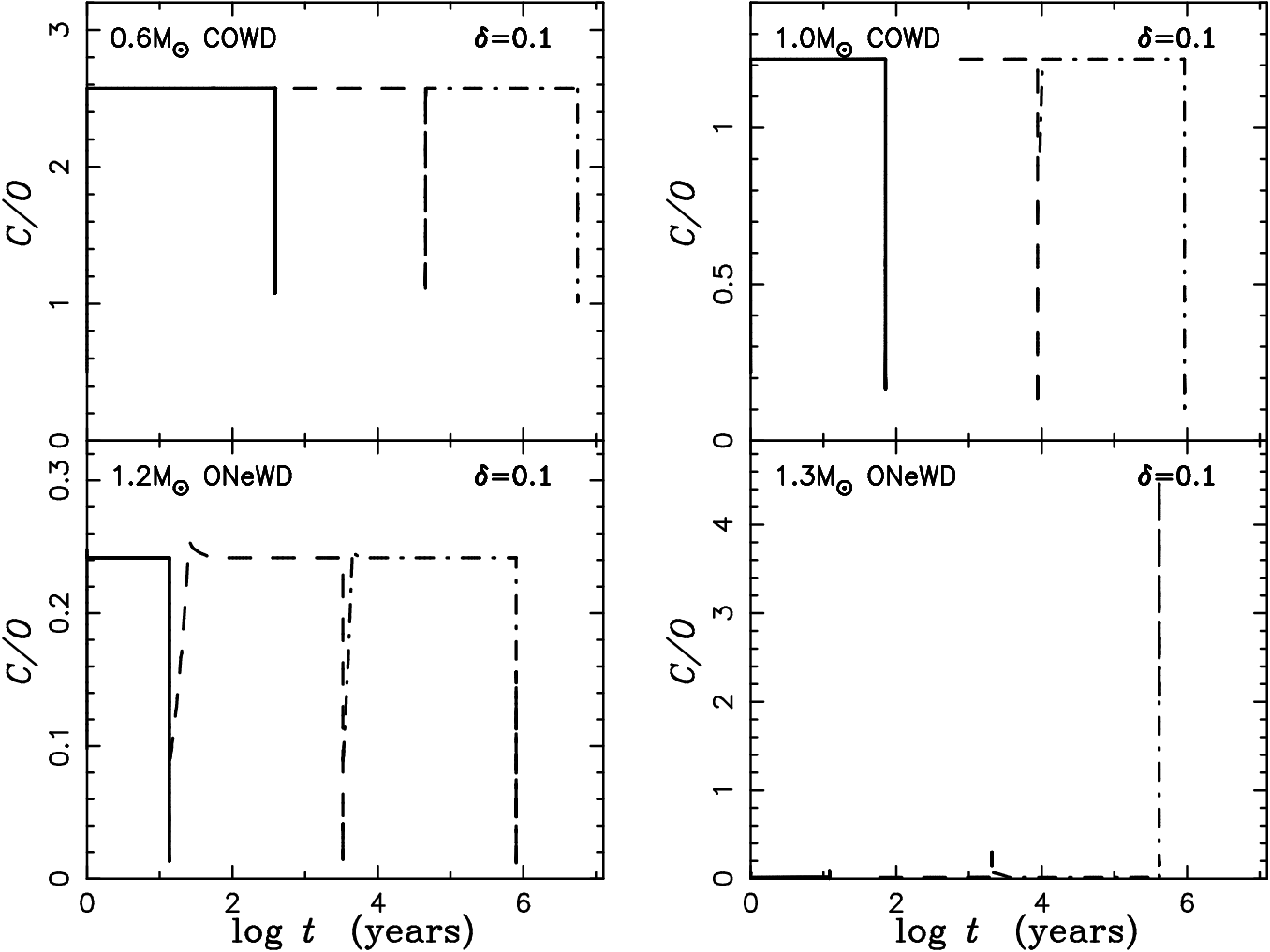}\\
\end{tabular}
\caption{The $C/O$ evolutions on the surface of WDs during nova eruptions.
The masses of WDs and the values of parameter $\delta$ are given in the top zone of every
panel. The solid, dashed and dash-dotted lines represent the models with the mass-accretion
rates of $1\times10^{-7}$, $1\times10^{-9}$ and $1\times10^{-11} \rm M_\odot$ yr$^{-1}$,
respectively.  }
\label{fig:coej}
\end{figure*}

Figure \ref{fig:coej} shows the $C/O$ evolutions for all nova models simulated in
the present paper. For the nova models with 0.6 $\rm M_\odot$ CO WD, $C/O$ of the
ejecta is always larger than 1.0 except the models with very low mass-accretion rate
($1\times10^{-11} \rm M_\odot$ yr$^{-1}$) and small mixing depth ($\delta=0.001$, and 0.01).
For the nova models with 1.0 $\rm M_\odot$ CO WD, $C/O$ can evolve from larger than 1.0 to
lower than 1.0 in all models, while the $C/O$ in the nova models with 1.2 $\rm M_\odot$ ONe WD
can do so only in the models with small mixing depth ($\delta=0.001$, and 0.01).
The main reason is that the mixing in these models with large mixing depth ($\delta=0.05$, and 0.1)
can result in a very low $C/O$ because the O abundance of the WD from the surface to the inside
quickly rises over C abundances (See Figure \ref{fig:wdch}). This reason is suitable to all models
with 1.3 $\rm M_\odot$ ONe WD. More interestingly, in the models with very low mass-accretion rate
($1\times10^{-11}\rm M_\odot$ yr$^{-1}$),
the TNR deletes amounts of O element so that $C/O$ in the ejecta changes from smaller than 1.0 to higher than 1.0.
This means that the silicate dust may be produced at first, and the amorphous carbon dust forms after that.

The bimodal dust species have been observed in following six novae:
V1370 Aql, V842 Cen, QV Vul, V2676 Oph, V1280 Sco and V1065 Cent
\citep{Strope2010,Helton2010,Sakon2016,Kawakita2017}.
Here, the masses of CO WDs in V1280 Sco, V842 Cen and V2676 Oph
have been observationally estimated. They are $\leq 0.6$, 0.88 and 0.6
(by slow evolution of light curves) or 1.1 (by nucleosynthesis) $\rm M_\odot$,
respectively \citep{Sakon2016,Luna2012,Kawakita2017}.
These observations have a weak constrain on our model.
%However, V1065 Cent is ONe WD. This means a very small mix depth, that is, the mixture only occurs
%in a thin layer close to the surface of WD, which is consistent with
%the results of \cite{Prialnik1984} and \cite{Alexakis2004}.

%\subsection{Dust Nucleation}
%As the last section discusses, it is possible that
%the bimodal dust species are produced in a nova eruption.
%However, based on the classical nucleation theory,
%the dust grains can be formed until the gas is supersaturated \citep{Becker1935,Feder1966}.
%Following \cite{Derdzinski2017}, we only consider the possibility for
%the formation of carbon grains and Mg$_2$SiO$_4$ which represents the silicate grain
%population during a nova eruption.

%In the classical nucleation theory, it is determined by the ration of the gas density
%to the equilibrium density ($n_{\rm eq}$) whether a gas becomes supersaturated.
%For carbon grains, this ratio is given by
%\begin{equation}
%S_{\rm C}=n_{\rm C}/n_{\rm eq},
%\label{eq:sc}
%\end{equation}
In addition,  based on \cite{Yaron2005}, the different core temperatures of WDs can result
in an uncertainty up to a factor of about 10. Considering that the input parameters
(the mixing depth, the WD's mass and the mass-accretion rate) in this work have lead to
a large scatter of our results, the present paper does not discuss its effects.
Simultaneously, we have not conducted 3D
simulations in this work since they are extremely time-
consuming and 1D simulations are accurate enough for our
goals.

\section{Conclusions}
In order to discuss the possibility for the formation of bimodal dust species,
we use MESA to investigate the chemical abundances of nova ejecta.
Having considered the chemical profiles of WD, the new input parameter,
mixing depth, is introduced to describe the mixing zone when TNR occurs.
The effects of WD mass, mass-accretion rate and mixing depth on the nova eruption
are studied.  The effects of the first two parameters ( WD mass and mass-accretion rate)
are similar with the previous work, that is, the lower mass-accretion
rate of a WD is, the stronger the TNR is, and the higher WD's mass is, the shorter and the stronger nova eruption is.
For new parameter---the mixing depth, we find that a large mixing depth can trigger an earlier TNR because it can provide more C and O elements for the
hydrogen burning in CNO cycle. Therefore, in these models, the hydrogen mass burned is less than that
with a small mixing depth and the maximum temperature during TNR is also lower.

We focus on the $C/O$ evolution during a whole nova ejecta, and find that it greatly
depends on the mixing depth. This means that the chemical profiles of WD greatly affect
nova eruption.
For the models of CO or ONe WDs with a small mixing depth ($\delta=0.001$), the $C/O$ of
nova ejecta may be larger than 1.0 at the
beginning of the nova eruption, and then quickly becomes
smaller than 1.0 due to the nucleosynthesises of TNR.
However, for a large mixing depth ($\delta\ge0.05$), the $C/O$ for ONe WD is always
smaller than 1.0, and even for model of ONe WD with low mass-accretion ($10^{-11}\rm M_\odot$yr$^{-1}$),
the $C/O$ of the ejecta changes from smaller than 1.0 to higher than
1.0.
Considering the bimodal dust species have been observed in CO  and ONe WDs,
we suggest that the mixing depth should be a very small value. This
means that the mixing only occurs in a thin layer close to the surface of WD.

\section*{Acknowledgments}
This work received the generous support of the  National Natural Science Foundation of China,
project Nos. 11763007, 11863005, 11803026
and 11503008. We would also like to express our gratitude to the Tianshan Youth Project of Xinjiang No.2017Q014.

\appendix
\section{Input Parameters and Yields of Nova Models}
Tables 1---4 show the input parameters for all models and results including the envelope's mass ($M_{\rm en}$) before TNR occurs,
the mass ejected ($M_{\rm ej}$) during nova eruption, and the maximum temperature ($T_{\rm max}$) of TNR during eruption.
The chemical abundances (mass fraction) of isotopes from $^1$H to $^{40}$Ca are given.

\begin{landscape}
\begin{table*}
  \caption{Input parameters for the models with $0.6 \rm M_\odot$ CO WD and results. $\Delta M_{\rm en}$ and $\Delta M_{\rm ej}$ are the envelope's mass
  before TNR occurs and the mass ejected during nova eruptions in unit of $10^{-5}\rm M_\odot$. $T_{\rm max}$ is
  the maximum temperature (in unit of $10^8 \rm K$) of TNR during nova eruption . $^{\rm i}X$ represents the yields (mass fraction)
  of isotope $X$ in the nova ejecta.   }
  \tabcolsep0.5mm
  \begin{tabular}{ccccccccccccc}
\cline{1-13}
\multicolumn{1}{|c|}{}&\multicolumn{12}{|c|}{$M_{\rm WD}=0.6\rm M_\odot$ } \\
\multicolumn{1}{|c|}{}&\multicolumn{3}{|c|}{$\delta=0.001$ }& \multicolumn{3}{|c|}{$\delta=0.01$ }&
\multicolumn{3}{|c|}{$\delta=0.05$ }&\multicolumn{3}{|c|}{$\delta=0.1$ }\\
\multicolumn{1}{|c|}{Models}&\multicolumn{1}{|c|}{$\dot{M}=10^{-7}$}&\multicolumn{1}{|c|}{$\dot{M}=10^{-9}$}
&\multicolumn{1}{|c|}{$\dot{M}=10^{-11}$}&\multicolumn{1}{|c|}{$\dot{M}=10^{-7}$}&\multicolumn{1}{|c|}{$\dot{M}=10^{-9}$}
&\multicolumn{1}{|c|}{$\dot{M}=10^{-11}$}&\multicolumn{1}{|c|}{$\dot{M}=10^{-7}$}&\multicolumn{1}{|c|}{$\dot{M}=10^{-9}$}
&\multicolumn{1}{|c|}{$\dot{M}=10^{-11}$}&\multicolumn{1}{|c|}{$\dot{M}=10^{-7}$}&\multicolumn{1}{|c|}{$\dot{M}=10^{-9}$}
&\multicolumn{1}{|c|}{$\dot{M}=10^{-11}$}\\

\cline{1-13} Results&1&2&3&4&5&6&7&8&9&10&11&12\\
$\Delta M_{\rm en}$&5.8&7.7&19.8&5.4&6.8&16.5&4.4&4.9&6.3&4.0&4.5&6.5\\
$\Delta M_{\rm ej}$&5.1&7.3&18.5&4.7&6.4&15.9&3.8&4.6&6.1&3.8&4.4&5.5\\
$T_{\rm max}$&1.0&1.1&1.3&1.0&1.0&1.0&0.9&1.0&1.0&0.9&1.0&1.0\\
$^{1}$H&$3.7\times10^{-1}$&$3.7\times10^{-1}$&$3.6\times10^{-1}$&
        $3.5\times10^{-1}$&$3.5\times10^{-1}$&$3.3\times10^{-1}$&
        $3.4\times10^{-1}$&$3.4\times10^{-1}$&$3.3\times10^{-1}$&
        $3.4\times10^{-1}$&$3.4\times10^{-1}$&$3.4\times10^{-1}$\\
$^{2}$H&$2.1\times10^{-10}$&$4.1\times10^{-12}$&$5.5\times10^{-13}$&
        $2.6\times10^{-10}$&$4.6\times10^{-11}$&$4.1\times10^{-12}$&
        $1.1\times10^{-9}$&$1.2\times10^{-10}$&$8.6\times10^{-11}$&
        $4.4\times10^{-10}$&$1.5\times10^{-10}$&$5.2\times10^{-11}$\\
$^{3}$He&$7.4\times10^{-6}$&$4.5\times10^{-6}$&$2.6\times10^{-7}$&
         $9.0\times10^{-6}$&$6.5\times10^{-6}$&$3.4\times10^{-7}$&
         $1.4\times10^{-5}$&$1.3\times10^{-5}$&$1.4\times10^{-5}$&
         $1.5\times10^{-5}$&$1.4\times10^{-5}$&$1.6\times10^{-5}$\\
$^{4}$He&$5.9\times10^{-1}$&$5.9\times10^{-1}$&$5.9\times10^{-1}$&
         $5.4\times10^{-1}$&$5.4\times10^{-1}$&$5.5\times10^{-1}$&
         $3.0\times10^{-1}$&$3.0\times10^{-1}$&$3.0\times10^{-1}$&
         $2.2\times10^{-1}$&$2.2\times10^{-1}$&$2.2\times10^{-1}$\\
%$^{6}$Li&$5.2\times10^{-15}$&$1.0\times10^{-19}$&$1.4\times10^{-17}$&
%         $6.6\times10^{-15}$&$1.2\times10^{-15}$&$3.3\times10^{-16}$&
%         $3.1\times10^{-14}$&$3.6\times10^{-15}$&$4.4\times10^{-15}$&
%         $1.4\times10^{-14}$&$4.9\times10^{-15}$&$3.7\times10^{-15}$\\
$^{7}$Li&$1.6\times10^{-9}$&$8.5\times10^{-10}$&$1.1\times10^{-10}$&
         $1.7\times10^{-9}$&$1.1\times10^{-9}$&$1.9\times10^{-9}$&
         $9.9\times10^{-10}$&$7.3\times10^{-10}$&$6.9\times10^{-10}$&
         $5.8\times10^{-10}$&$4.5\times10^{-10}$&$4.3\times10^{-10}$\\
$^{7}$Be&$2.0\times10^{-5}$&$1.5\times10^{-5}$&$1.3\times10^{-5}$&
         $1.9\times10^{-5}$&$1.7\times10^{-5}$&$1.3\times10^{-5}$&
         $1.1\times10^{-5}$&$9.6\times10^{-6}$&$1.2\times10^{-5}$&
         $7.4\times10^{-6}$&$6.6\times10^{-6}$&$8.2\times10^{-6}$\\
%$^{8}$Be&---&---&---&---&---&---&---&---&---&---&---&---\\
%$^{10}$B&$7.6\times10^{-15}$&$1.6\times10^{-19}$&$2.0\times10^{-17}$&
%         $9.6\times10^{-15}$&$1.8\times10^{-15}$&$4.9\times10^{-16}$&
%         $4.5\times10^{-14}$&$5.3\times10^{-15}$&$6.5\times10^{-15}$&
%         $2.1\times10^{-14}$&$7.2\times10^{-15}$&$5.4\times10^{-15}$\\
%$^{11}$B&$3.4\times10^{-14}$&$7.4\times10^{-19}$&$9.1\times10^{-17}$&
%         $4.3\times10^{-14}$&$8.0\times10^{-15}$&$2.2\times10^{-15}$&
 %        $2.0\times10^{-13}$&$2.4\times10^{-14}$&$2.9\times10^{-14}$&
 %        $9.6\times10^{-14}$&$3.2\times10^{-14}$&$2.4\times10^{-14}$\\
$^{12}$C&$9.8\times10^{-3}$&$7.4\times10^{-3}$&$1.4\times10^{-3}$&
         $3.9\times10^{-2}$&$3.3\times10^{-2}$&$4.4\times10^{-3}$&
         $1.2\times10^{-1}$&$1.1\times10^{-1}$&$9.8\times10^{-2}$&
         $1.2\times10^{-1}$&$1.3\times10^{-1}$&$1.1\times10^{-1}$\\
$^{13}$C&$3.5\times10^{-3}$&$2.9\times10^{-3}$&$5.2\times10^{-4}$&
         $1.4\times10^{-2}$&$1.2\times10^{-2}$&$1.4\times10^{-3}$&
         $3.8\times10^{-2}$&$3.8\times10^{-2}$&$3.6\times10^{-2}$&
         $4.0\times10^{-2}$&$4.1\times10^{-2}$&$4.0\times10^{-2}$\\
$^{14}$N&$1.7\times10^{-2}$&$2.1\times10^{-2}$&$3.1\times10^{-2}$&
         $4.2\times10^{-2}$&$5.1\times10^{-2}$&$9.5\times10^{-2}$&
         $9.7\times10^{-2}$&$9.9\times10^{-2}$&$1.2\times10^{-1}$&
         $1.1\times10^{-1}$&$1.1\times10^{-1}$&$1.3\times10^{-1}$\\
$^{15}$N&$5.0\times10^{-7}$&$5.9\times10^{-7}$&$8.9\times10^{-7}$&
         $1.2\times10^{-6}$&$1.4\times10^{-6}$&$2.8\times10^{-6}$&
         $2.6\times10^{-6}$&$2.7\times10^{-6}$&$3.3\times10^{-6}$&
         $3.1\times10^{-6}$&$3.0\times10^{-6}$&$3.5\times10^{-6}$\\
$^{16}$O&$5.2\times10^{-4}$&$5.1\times10^{-3}$&$5.4\times10^{-3}$&
         $6.9\times10^{-3}$&$7.1\times10^{-3}$&$6.2\times10^{-3}$&
         $8.9\times10^{-2}$&$8.8\times10^{-2}$&$8.8\times10^{-2}$&
         $1.5\times10^{-1}$&$1.5\times10^{-1}$&$1.4\times10^{-1}$\\
$^{17}$O&$1.1\times10^{-5}$&$1.4\times10^{-5}$&$4.2\times10^{-5}$&
         $1.4\times10^{-5}$&$1.8\times10^{-5}$&$1.6\times10^{-5}$&
         $1.5\times10^{-4}$&$1.6\times10^{-4}$&$1.9\times10^{-4}$&
         $2.5\times10^{-4}$&$2.5\times10^{-4}$&$2.9\times10^{-4}$\\
$^{18}$O&$4.4\times10^{-9}$&$5.0\times10^{-9}$&$1.6\times10^{-8}$&
         $5.8\times10^{-9}$&$6.1\times10^{-9}$&$9.8\times10^{-9}$&
         $5.9\times10^{-8}$&$5.7\times10^{-8}$&$6.2\times10^{-8}$&
         $9.5\times10^{-8}$&$8.9\times10^{-8}$&$9.4\times10^{-8}$\\
%$^{18}$F&---&---&$1.8\times10^{-12}$&---&$1.5\times10^{-33}$&---&$1.5\times10^{-25}$&
%         $7.6\times10^{-23}$&$5.7\times10^{-20}$&$7.1\times10^{-24}$&
%         $5.4\times10^{-22}$&$1.6\times10^{-19}$\\
$^{19}$F&$2.8\times10^{-7}$&$2.4\times10^{-7}$&$3.2\times10^{-8}$&
         $4.4\times10^{-7}$&$3.9\times10^{-7}$&$9.9\times10^{-8}$&
         $3.9\times10^{-7}$&$3.8\times10^{-7}$&$3.5\times10^{-7}$&
         $3.7\times10^{-7}$&$3.7\times10^{-7}$&$3.5\times10^{-7}$\\
$^{20}$Ne&$1.9\times10^{-3}$&$1.9\times10^{-3}$&$1.9\times10^{-3}$&
          $1.7\times10^{-3}$&$1.7\times10^{-3}$&$1.8\times10^{-3}$&
          $1.7\times10^{-3}$&$1.7\times10^{-3}$&$1.7\times10^{-3}$&
          $1.6\times10^{-3}$&$1.6\times10^{-3}$&$1.7\times10^{-3}$\\
$^{21}$Ne&$5.2\times10^{-6}$&$5.2\times10^{-6}$&$5.2\times10^{-6}$&
          $4.3\times10^{-6}$&$4.3\times10^{-6}$&$4.3\times10^{-6}$&
          $4.1\times10^{-6}$&$4.2\times10^{-6}$&$4.2\times10^{-6}$&
          $4.0\times10^{-6}$&$4.1\times10^{-6}$&$4.1\times10^{-6}$\\
$^{22}$Ne&$2.2\times10^{-3}$&$2.2\times10^{-3}$&$1.5\times10^{-3}$&
          $8.2\times10^{-3}$&$8.0\times10^{-3}$&$4.5\times10^{-3}$&
          $8.4\times10^{-3}$&$8.4\times10^{-3}$&$8.3\times10^{-3}$&
          $8.0\times10^{-3}$&$8.1\times10^{-3}$&$8.2\times10^{-3}$\\
$^{23}$Na&$2.4\times10^{-4}$&$3.1\times10^{-4}$&$9.3\times10^{-4}$&
          $7.3\times10^{-4}$&$9.0\times10^{-4}$&$4.6\times10^{-3}$&
          $6.6\times10^{-4}$&$6.8\times10^{-4}$&$7.8\times10^{-4}$&
          $7.1\times10^{-4}$&$6.9\times10^{-4}$&$7.6\times10^{-4}$\\
$^{24}$Mg&$1.2\times10^{-3}$&$1.2\times10^{-3}$&$1.2\times10^{-3}$&
          $1.9\times10^{-3}$&$1.9\times10^{-3}$&$1.9\times10^{-3}$&
          $1.8\times10^{-3}$&$1.8\times10^{-3}$&$1.8\times10^{-3}$&
          $1.7\times10^{-3}$&$1.8\times10^{-3}$&$1.8\times10^{-3}$\\
$^{25}$Mg&$8.6\times10^{-5}$&$8.6\times10^{-5}$&$8.6\times10^{-5}$&
          $7.2\times10^{-5}$&$7.2\times10^{-5}$&$7.2\times10^{-5}$&
          $6.9\times10^{-5}$&$6.9\times10^{-5}$&$6.9\times10^{-5}$&
          $6.7\times10^{-5}$&$6.7\times10^{-5}$&$6.8\times10^{-5}$\\
$^{26}$Mg&$1.2\times10^{-4}$&$1.2\times10^{-4}$&$1.2\times10^{-4}$&
          $8.4\times10^{-5}$&$8.4\times10^{-5}$&$8.3\times10^{-5}$&
          $7.9\times10^{-5}$&$7.9\times10^{-5}$&$7.9\times10^{-5}$&
          $7.8\times10^{-5}$&$7.8\times10^{-5}$&$7.8\times10^{-5}$\\
$^{27}$Al&$7.5\times10^{-5}$&$7.5\times10^{-5}$&$7.5\times10^{-5}$&
          $6.3\times10^{-5}$&$6.3\times10^{-5}$&$6.4\times10^{-5}$&
          $6.0\times10^{-5}$&$6.0\times10^{-5}$&$6.0\times10^{-5}$&
          $5.8\times10^{-5}$&$5.9\times10^{-5}$&$5.9\times10^{-5}$\\
$^{28}$Si&$8.5\times10^{-4}$&$8.5\times10^{-4}$&$8.4\times10^{-4}$&
          $7.1\times10^{-4}$&$7.1\times10^{-4}$&$7.1\times10^{-4}$&
          $6.8\times10^{-4}$&$6.8\times10^{-4}$&$6.8\times10^{-4}$&
          $6.6\times10^{-4}$&$6.6\times10^{-4}$&$6.7\times10^{-4}$\\
$^{29}$Si&$4.4\times10^{-5}$&$4.4\times10^{-5}$&$4.4\times10^{-5}$&
          $3.7\times10^{-5}$&$3.7\times10^{-5}$&$3.7\times10^{-5}$&
          $3.5\times10^{-5}$&$3.6\times10^{-5}$&$3.6\times10^{-5}$&
          $3.5\times10^{-5}$&$3.5\times10^{-5}$&$3.5\times10^{-5}$\\
$^{30}$Si&$3.0\times10^{-5}$&$3.0\times10^{-5}$&$3.0\times10^{-5}$&
          $2.5\times10^{-5}$&$2.5\times10^{-5}$&$2.5\times10^{-5}$&
          $2.4\times10^{-5}$&$2.4\times10^{-5}$&$2.4\times10^{-5}$&
          $2.4\times10^{-5}$&$2.4\times10^{-5}$&$2.4\times10^{-5}$\\
$^{31}$P&$9.7\times10^{-6}$&$9.7\times10^{-6}$&$9.7\times10^{-6}$&
         $7.9\times10^{-6}$&$7.9\times10^{-6}$&$7.9\times10^{-6}$&
         $7.6\times10^{-6}$&$7.6\times10^{-6}$&$7.6\times10^{-6}$&
         $7.4\times10^{-6}$&$7.4\times10^{-6}$&$7.5\times10^{-6}$\\
$^{32}$S&$4.7\times10^{-4}$&$4.7\times10^{-4}$&$4.7\times10^{-4}$&
         $3.9\times10^{-4}$&$3.9\times10^{-4}$&$3.9\times10^{-4}$&
         $3.7\times10^{-4}$&$3.7\times10^{-4}$&$3.7\times10^{-4}$&
         $3.6\times10^{-4}$&$3.6\times10^{-4}$&$3.7\times10^{-4}$\\
$^{33}$S&$3.9\times10^{-6}$&$3.9\times10^{-6}$&$3.9\times10^{-6}$&
         $3.2\times10^{-6}$&$3.2\times10^{-6}$&$3.2\times10^{-6}$&
         $3.0\times10^{-6}$&$3.0\times10^{-6}$&$3.0\times10^{-6}$&
         $2.9\times10^{-6}$&$3.0\times10^{-6}$&$3.0\times10^{-6}$\\
$^{34}$S&$2.2\times10^{-5}$&$2.2\times10^{-5}$&$2.2\times10^{-5}$&
         $1.8\times10^{-5}$&$1.8\times10^{-5}$&$1.8\times10^{-5}$&
         $1.8\times10^{-5}$&$1.8\times10^{-5}$&$1.8\times10^{-5}$&
         $1.7\times10^{-5}$&$1.7\times10^{-5}$&$1.7\times10^{-5}$\\
$^{35}$Cl&$3.8\times10^{-6}$&$3.8\times10^{-6}$&$3.8\times10^{-6}$&
          $3.3\times10^{-6}$&$3.3\times10^{-6}$&$3.3\times10^{-6}$&
          $3.1\times10^{-6}$&$3.1\times10^{-6}$&$3.1\times10^{-6}$&
          $3.0\times10^{-6}$&$3.1\times10^{-6}$&$3.1\times10^{-6}$\\
%$^{36}$Cl&$2.9\times10^{-18}$&$5.6\times10^{-18}$&$5.0\times10^{-17}$&
%          $7.6\times10^{-18}$&$1.3\times10^{-17}$&$2.6\times10^{-17}$&
%          $7.8\times10^{-18}$&$9.0\times10^{-18}$&$1.4\times10^{-17}$&
%          $7.8\times10^{-18}$&$6.4\times10^{-18}$&$9.7\times10^{-18}$\\
$^{37}$Cl&$6.4\times10^{-7}$&$6.3\times10^{-7}$&$6.4\times10^{-7}$&
          $6.3\times10^{-7}$&$6.4\times10^{-7}$&$6.3\times10^{-7}$&
          $6.5\times10^{-7}$&$6.5\times10^{-7}$&$6.5\times10^{-7}$&
          $6.6\times10^{-7}$&$6.5\times10^{-7}$&$6.5\times10^{-7}$\\
$^{36}$Ar&$3.9\times10^{-5}$&$3.9\times10^{-5}$&$3.9\times10^{-5}$&
          $3.9\times10^{-5}$&$3.9\times10^{-5}$&$3.9\times10^{-5}$&
          $4.1\times10^{-5}$&$4.0\times10^{-5}$&$4.1\times10^{-5}$&
          $4.2\times10^{-5}$&$4.1\times10^{-5}$&$4.1\times10^{-5}$\\
%$^{37}$Ar&$9.3\times10^{-18}$&$1.7\times10^{-17}$&$1.5\times10^{-16}$&
%          $2.9\times10^{-17}$&$4.5\times10^{-17}$&$5.8\times10^{-17}$&
%          $3.8\times10^{-17}$&$4.2\times10^{-17}$&$7.0\times10^{-17}$&
%          $4.0\times10^{-17}$&$3.4\times10^{-17}$&$5.3\times10^{-17}$\\
$^{38}$Ar&$7.7\times10^{-5}$&$7.7\times10^{-5}$&$7.7\times10^{-5}$&
          $5.8\times10^{-5}$&$5.8\times10^{-5}$&$5.8\times10^{-5}$&
          $5.2\times10^{-5}$&$5.2\times10^{-5}$&$5.2\times10^{-5}$&
          $4.8\times10^{-5}$&$4.9\times10^{-5}$&$5.0\times10^{-5}$\\
%$^{38}$K&---&---&$3.5\times10^{-19}$&$2.7\times10^{-17}$&---&---&---&---&---&
%          $4.0\times10^{-31}$&---&$3.5\times10^{-21}$\\
$^{39}$K&$4.5\times10^{-6}$&$4.5\times10^{-6}$&$4.5\times10^{-6}$&
         $3.8\times10^{-6}$&$3.7\times10^{-6}$&$3.8\times10^{-6}$&
         $3.6\times10^{-6}$&$3.6\times10^{-6}$&$3.6\times10^{-6}$&
         $3.5\times10^{-6}$&$3.5\times10^{-6}$&$3.5\times10^{-6}$\\
$^{40}$Ca&$7.7\times10^{-5}$&$7.7\times10^{-5}$&$7.7\times10^{-5}$&
          $6.4\times10^{-5}$&$6.4\times10^{-5}$&$6.4\times10^{-5}$&
          $5.9\times10^{-5}$&$5.9\times10^{-5}$&$5.9\times10^{-5}$&
          $5.5\times10^{-5}$&$5.7\times10^{-5}$&$5.7\times10^{-5}$\\
%$^{41}$Ca&$2.0\times10^{-17}$&$5.9\times10^{-17}$&$5.8\times10^{-16}$&
%          $5.9\times10^{-17}$&$1.3\times10^{-16}$&$2.8\times10^{-16}$&
%          $6.6\times10^{-17}$&$8.9\times10^{-17}$&$1.5\times10^{-16}$&
%          $6.7\times10^{-17}$&$6.1\times10^{-17}$&$9.7\times10^{-17}$\\
\hline
 \label{tab:0.6}
\end{tabular}
\end{table*}
\end{landscape}

\begin{landscape}
\begin{table*}
  \caption{Similar with Table \ref{tab:0.6} but for models with 1.0 $\rm M_\odot$ CO WD.}
  \tabcolsep0.5mm
  \begin{tabular}{ccccccccccccc}
\cline{1-13}
\multicolumn{1}{|c|}{}&\multicolumn{12}{|c|}{$M_{\rm WD}=1.0M_\odot$ } \\
\multicolumn{1}{|c|}{}&\multicolumn{3}{|c|}{$\delta=0.001$ }& \multicolumn{3}{|c|}{$\delta=0.01$ }&
\multicolumn{3}{|c|}{$\delta=0.05$ }&\multicolumn{3}{|c|}{$\delta=0.1$ }\\
\multicolumn{1}{|c|}{Models}&\multicolumn{1}{|c|}{$\dot{M}=10^{-7}$}&\multicolumn{1}{|c|}{$\dot{M}=10^{-9}$}
&\multicolumn{1}{|c|}{$\dot{M}=10^{-11}$}&\multicolumn{1}{|c|}{$\dot{M}=10^{-7}$}&\multicolumn{1}{|c|}{$\dot{M}=10^{-9}$}
&\multicolumn{1}{|c|}{$\dot{M}=10^{-11}$}&\multicolumn{1}{|c|}{$\dot{M}=10^{-7}$}&\multicolumn{1}{|c|}{$\dot{M}=10^{-9}$}
&\multicolumn{1}{|c|}{$\dot{M}=10^{-11}$}&\multicolumn{1}{|c|}{$\dot{M}=10^{-7}$}&\multicolumn{1}{|c|}{$\dot{M}=10^{-9}$}
&\multicolumn{1}{|c|}{$\dot{M}=10^{-11}$}\\

\cline{1-13} Results&1&2&3&4&5&6&7&8&9&10&11&12\\
$\Delta M_{\rm en}$&1.4&2.3&4.4&0.7&0.9&1.0&0.7&0.9&0.9&0.7&0.9&0.9\\
$\Delta M_{\rm ej}$&1.0&1.9&3.8&0.7&0.9&0.9&0.7&0.8&0.9&0.7&0.8&0.9\\
$T_{\rm max}$&1.5&1.7&1.8&1.3&1.3&1.4&1.3&1.3&1.3&1.3&1.3&1.3\\
$^{1}$H&$3.4\times10^{-1}$&$3.3\times10^{-1}$&$3.2\times10^{-1}$&
        $3.2\times10^{-1}$&$3.1\times10^{-1}$&$3.1\times10^{-1}$&
        $3.2\times10^{-1}$&$3.2\times10^{-1}$&$3.1\times10^{-1}$&
        $3.2\times10^{-1}$&$3.2\times10^{-1}$&$3.1\times10^{-1}$\\
$^{2}$H&$3.8\times10^{-10}$&$1.5\times10^{-11}$&$3.2\times10^{-11}$&
        $4.5\times10^{-11}$&$2.9\times10^{-11}$&$3.9\times10^{-12}$&
        $2.2\times10^{-10}$&$3.0\times10^{-11}$&$3.7\times10^{-12}$&
        $6.4\times10^{-11}$&$1.2\times10^{-11}$&$8.7\times10^{-12}$\\
$^{3}$He&$1.3\times10^{-9}$&$5.4\times10^{-11}$&$1.1\times10^{-10}$&
         $7.0\times10^{-6}$&$4.9\times10^{-6}$&$5.4\times10^{-6}$&
         $7.2\times10^{-6}$&$6.3\times10^{-6}$&$5.5\times10^{-6}$&
         $7.5\times10^{-6}$&$6.3\times10^{-6}$&$4.5\times10^{-6}$\\
$^{4}$He&$6.3\times10^{-1}$&$6.4\times10^{-1}$&$6.6\times10^{-1}$&
         $2.1\times10^{-1}$&$2.1\times10^{-1}$&$2.1\times10^{-1}$&
         $1.5\times10^{-1}$&$1.6\times10^{-1}$&$1.6\times10^{-1}$&
         $1.5\times10^{-1}$&$1.5\times10^{-1}$&$1.6\times10^{-1}$\\
%$^{6}$Li&$9.7\times10^{-15}$&$3.9\times10^{-16}$&$8.7\times10^{-16}$&
%         $1.5\times10^{-15}$&$8.5\times10^{-16}$&$3.2\times10^{-16}$&
%         $7.5\times10^{-15}$&$9.0\times10^{-16}$&$3.5\times10^{-16}$&
%         $2.2\times10^{-15}$&$4.2\times10^{-16}$&$6.3\times10^{-16}$\\
$^{7}$Li&$8.2\times10^{-10}$&$3.3\times10^{-10}$&$6.7\times10^{-11}$&
         $1.3\times10^{-10}$&$1.0\times10^{-10}$&$9.4\times10^{-11}$&
         $1.1\times10^{-10}$&$8.1\times10^{-11}$&$7.9\times10^{-11}$&
         $1.0\times10^{-10}$&$8.1\times10^{-11}$&$8.7\times10^{-11}$\\
$^{7}$Be&$2.2\times10^{-5}$&$1.3\times10^{-5}$&$5.7\times10^{-6}$&
         $1.4\times10^{-5}$&$1.3\times10^{-5}$&$1.2\times10^{-5}$&
         $1.1\times10^{-5}$&$1.0\times10^{-5}$&$9.9\times10^{-6}$&
         $1.0\times10^{-5}$&$9.7\times10^{-6}$&$9.9\times10^{-6}$\\
%$^{8}$Be&---&---&---&---&---&---&---&---&---&---&---&---\\
%$^{10}$B&$1.4\times10^{-14}$&$5.7\times10^{-16}$&$1.3\times10^{-15}$&
%         $2.2\times10^{-15}$&$1.2\times10^{-15}$&$4.8\times10^{-16}$&
%         $1.1\times10^{-14}$&$1.3\times10^{-15}$&$5.2\times10^{-16}$&
%         $3.3\times10^{-15}$&$6.3\times10^{-16}$&$9.2\times10^{-16}$\\
%$^{11}$B&$6.4\times10^{-14}$&$2.6\times10^{-15}$&$5.7\times10^{-15}$&
%         $9.8\times10^{-15}$&$5.6\times10^{-15}$&$2.2\times10^{-15}$&
%         $5.0\times10^{-14}$&$5.9\times10^{-15}$&$2.3\times10^{-15}$&
%         $1.5\times10^{-14}$&$2.8\times10^{-15}$&$4.1\times10^{-15}$\\
$^{12}$C&$4.6\times10^{-4}$&$4.9\times10^{-4}$&$5.5\times10^{-4}$&
         $3.6\times10^{-2}$&$2.5\times10^{-2}$&$2.4\times10^{-2}$&
         $3.2\times10^{-2}$&$2.6\times10^{-2}$&$2.1\times10^{-2}$&
         $3.3\times10^{-2}$&$2.5\times10^{-2}$&$1.7\times10^{-2}$\\
$^{13}$C&$1.4\times10^{-4}$&$1.5\times10^{-4}$&$1.7\times10^{-4}$&
         $1.6\times10^{-2}$&$1.1\times10^{-2}$&$1.2\times10^{-2}$&
         $1.4\times10^{-2}$&$1.2\times10^{-2}$&$9.2\times10^{-3}$&
         $1.4\times10^{-2}$&$1.2\times10^{-2}$&$6.6\times10^{-3}$\\
$^{14}$N&$1.3\times10^{-2}$&$1.3\times10^{-2}$&$1.4\times10^{-2}$&
         $2.2\times10^{-1}$&$2.4\times10^{-1}$&$2.4\times10^{-1}$&
         $2.3\times10^{-1}$&$2.4\times10^{-1}$&$2.5\times10^{-1}$&
         $2.2\times10^{-1}$&$2.3\times10^{-1}$&$2.5\times10^{-1}$\\
$^{15}$N&$1.6\times10^{-6}$&$2.1\times10^{-6}$&$3.9\times10^{-6}$&
         $6.2\times10^{-6}$&$8.2\times10^{-6}$&$7.7\times10^{-6}$&
         $6.5\times10^{-6}$&$7.4\times10^{-6}$&$8.1\times10^{-6}$&
         $6.2\times10^{-6}$&$7.3\times10^{-6}$&$8.9\times10^{-6}$\\
$^{16}$O&$3.0\times10^{-3}$&$2.2\times10^{-3}$&$1.0\times10^{-3}$&
         $1.8\times10^{-1}$&$1.8\times10^{-1}$&$1.8\times10^{-1}$&
         $2.4\times10^{-1}$&$2.4\times10^{-1}$&$2.4\times10^{-1}$&
         $2.5\times10^{-1}$&$2.5\times10^{-1}$&$2.5\times10^{-1}$\\
$^{17}$O&$1.7\times10^{-5}$&$1.4\times10^{-5}$&$9.1\times10^{-6}$&
         $1.2\times10^{-3}$&$1.4\times10^{-3}$&$1.6\times10^{-3}$&
         $1.6\times10^{-3}$&$1.8\times10^{-3}$&$2.0\times10^{-3}$&
         $1.6\times10^{-3}$&$1.9\times10^{-3}$&$1.9\times10^{-3}$\\
$^{18}$O&$1.9\times10^{-7}$&$2.3\times10^{-7}$&$1.7\times10^{-7}$&
         $3.9\times10^{-7}$&$5.3\times10^{-7}$&$6.7\times10^{-7}$&
         $4.7\times10^{-7}$&$6.7\times10^{-7}$&$8.0\times10^{-7}$&
         $5.1\times10^{-7}$&$7.1\times10^{-7}$&$8.1\times10^{-7}$\\
%$^{18}$F&$8.1\times10^{-27}$&$1.7\times10^{-22}$&$2.5\times10^{-14}$&
%         $5.0\times10^{-9}$&$6.8\times10^{-9}$&$7.2\times10^{-9}$&
%         $7.0\times10^{-9}$&$9.2\times10^{-9}$&$9.8\times10^{-9}$&
%         $6.9\times10^{-9}$&$9.3\times10^{-9}$&$1.1\times10^{-8}$\\
$^{19}$F&$4.4\times10^{-9}$&$3.4\times10^{-9}$&$2.1\times10^{-9}$&
         $3.4\times10^{-7}$&$4.9\times10^{-7}$&$5.1\times10^{-7}$&
         $4.0\times10^{-7}$&$5.7\times10^{-7}$&$6.8\times10^{-7}$&
         $4.0\times10^{-7}$&$5.9\times10^{-7}$&$6.9\times10^{-7}$\\
$^{20}$Ne&$2.2\times10^{-3}$&$2.2\times10^{-3}$&$2.1\times10^{-3}$&
          $1.9\times10^{-3}$&$1.9\times10^{-3}$&$1.9\times10^{-3}$&
          $1.9\times10^{-3}$&$1.9\times10^{-3}$&$2.0\times10^{-3}$&
          $1.9\times10^{-3}$&$1.9\times10^{-3}$&$1.9\times10^{-3}$\\
$^{21}$Ne&$4.3\times10^{-6}$&$4.3\times10^{-6}$&$4.3\times10^{-6}$&
          $4.3\times10^{-6}$&$4.1\times10^{-6}$&$4.3\times10^{-6}$&
          $4.2\times10^{-6}$&$4.2\times10^{-6}$&$4.2\times10^{-6}$&
          $4.2\times10^{-6}$&$4.2\times10^{-6}$&$4.1\times10^{-6}$\\
$^{22}$Ne&$3.7\times10^{-6}$&$1.7\times10^{-7}$&$8.8\times10^{-9}$&
          $6.8\times10^{-3}$&$6.0\times10^{-3}$&$6.4\times10^{-3}$&
          $7.1\times10^{-3}$&$6.9\times10^{-3}$&$6.6\times10^{-3}$&
          $7.2\times10^{-3}$&$6.9\times10^{-3}$&$6.1\times10^{-3}$\\
$^{23}$Na&$1.4\times10^{-4}$&$3.0\times10^{-6}$&$2.5\times10^{-8}$&
          $2.0\times10^{-3}$&$2.4\times10^{-3}$&$2.2\times10^{-3}$&
          $2.2\times10^{-3}$&$2.4\times10^{-3}$&$2.6\times10^{-3}$&
          $7.1\times10^{-4}$&$2.4\times10^{-3}$&$2.9\times10^{-3}$\\
$^{24}$Mg&$1.8\times10^{-3}$&$2.0\times10^{-3}$&$2.1\times10^{-3}$&
          $2.0\times10^{-3}$&$1.9\times10^{-3}$&$2.0\times10^{-3}$&
          $2.0\times10^{-3}$&$2.0\times10^{-3}$&$2.0\times10^{-3}$&
          $2.0\times10^{-3}$&$2.0\times10^{-3}$&$1.9\times10^{-3}$\\
$^{25}$Mg&$7.1\times10^{-5}$&$7.1\times10^{-5}$&$7.1\times10^{-5}$&
          $7.0\times10^{-5}$&$6.8\times10^{-5}$&$7.0\times10^{-5}$&
          $7.0\times10^{-5}$&$6.9\times10^{-5}$&$7.0\times10^{-5}$&
          $7.0\times10^{-5}$&$6.9\times10^{-5}$&$6.8\times10^{-5}$\\
$^{26}$Mg&$6.3\times10^{-5}$&$2.6\times10^{-5}$&$1.2\times10^{-8}$&
          $7.9\times10^{-5}$&$7.6\times10^{-5}$&$7.9\times10^{-5}$&
          $7.8\times10^{-5}$&$7.8\times10^{-5}$&$7.8\times10^{-5}$&
          $7.9\times10^{-5}$&$7.8\times10^{-5}$&$7.6\times10^{-5}$\\
$^{27}$Al&$3.5\times10^{-5}$&$7.2\times10^{-6}$&$3.9\times10^{-9}$&
          $6.1\times10^{-5}$&$5.7\times10^{-5}$&$5.9\times10^{-5}$&
          $6.0\times10^{-5}$&$5.9\times10^{-5}$&$5.8\times10^{-5}$&
          $6.0\times10^{-5}$&$5.9\times10^{-5}$&$5.7\times10^{-5}$\\
$^{28}$Si&$7.0\times10^{-4}$&$7.1\times10^{-4}$&$7.1\times10^{-4}$&
          $6.9\times10^{-4}$&$6.7\times10^{-4}$&$6.9\times10^{-4}$&
          $6.9\times10^{-4}$&$6.8\times10^{-4}$&$6.9\times10^{-4}$&
          $6.9\times10^{-4}$&$6.8\times10^{-4}$&$6.7\times10^{-4}$\\
$^{29}$Si&$3.7\times10^{-5}$&$3.7\times10^{-5}$&$3.7\times10^{-5}$&
          $3.6\times10^{-5}$&$3.5\times10^{-5}$&$3.6\times10^{-5}$&
          $3.6\times10^{-5}$&$3.6\times10^{-5}$&$3.6\times10^{-5}$&
          $3.6\times10^{-5}$&$3.6\times10^{-5}$&$3.5\times10^{-5}$\\
$^{30}$Si&$2.5\times10^{-5}$&$2.5\times10^{-5}$&$2.5\times10^{-5}$&
          $2.5\times10^{-5}$&$2.4\times10^{-5}$&$2.5\times10^{-5}$&
          $2.5\times10^{-5}$&$2.5\times10^{-5}$&$2.5\times10^{-5}$&
          $2.5\times10^{-5}$&$2.5\times10^{-5}$&$2.4\times10^{-5}$\\
$^{31}$P&$7.8\times10^{-6}$&$7.6\times10^{-6}$&$6.8\times10^{-6}$&
         $7.8\times10^{-6}$&$7.4\times10^{-6}$&$7.8\times10^{-6}$&
         $7.7\times10^{-6}$&$7.6\times10^{-6}$&$7.7\times10^{-6}$&
         $7.7\times10^{-6}$&$7.6\times10^{-6}$&$7.5\times10^{-6}$\\
$^{32}$S&$3.8\times10^{-4}$&$3.8\times10^{-4}$&$3.8\times10^{-4}$&
         $3.8\times10^{-4}$&$3.7\times10^{-4}$&$3.8\times10^{-4}$&
         $3.8\times10^{-4}$&$3.7\times10^{-4}$&$3.8\times10^{-4}$&
         $3.8\times10^{-4}$&$3.8\times10^{-4}$&$3.7\times10^{-4}$\\
$^{33}$S&$3.1\times10^{-6}$&$3.1\times10^{-6}$&$3.1\times10^{-6}$&
         $3.1\times10^{-6}$&$3.0\times10^{-6}$&$3.1\times10^{-6}$&
         $3.1\times10^{-6}$&$3.1\times10^{-6}$&$3.1\times10^{-6}$&
         $3.1\times10^{-6}$&$3.1\times10^{-6}$&$3.0\times10^{-6}$\\
$^{34}$S&$1.8\times10^{-5}$&$1.8\times10^{-5}$&$1.7\times10^{-5}$&
         $1.8\times10^{-5}$&$1.7\times10^{-5}$&$1.8\times10^{-5}$&
         $1.8\times10^{-5}$&$1.8\times10^{-5}$&$1.8\times10^{-5}$&
         $1.8\times10^{-5}$&$1.8\times10^{-5}$&$1.7\times10^{-5}$\\
$^{35}$Cl&$3.2\times10^{-6}$&$3.3\times10^{-6}$&$4.0\times10^{-6}$&
          $3.2\times10^{-6}$&$3.1\times10^{-6}$&$3.2\times10^{-6}$&
          $3.2\times10^{-6}$&$3.2\times10^{-6}$&$3.2\times10^{-6}$&
          $3.2\times10^{-6}$&$3.2\times10^{-6}$&$3.1\times10^{-6}$\\
%$^{36}$Cl&$7.3\times10^{-16}$&$4.5\times10^{-15}$&$1.1\times10^{-14}$&
%          $1.6\times10^{-16}$&$2.1\times10^{-16}$&$2.3\times10^{-16}$&
%          $1.4\times10^{-16}$&$1.6\times10^{-16}$&$1.9\times10^{-16}$&
%          $1.2\times10^{-16}$&$1.6\times10^{-16}$&$1.8\times10^{-16}$\\
$^{37}$Cl&$6.3\times10^{-7}$&$6.3\times10^{-7}$&$6.3\times10^{-7}$&
          $6.3\times10^{-7}$&$6.3\times10^{-7}$&$6.3\times10^{-7}$&
          $6.3\times10^{-7}$&$6.3\times10^{-7}$&$6.3\times10^{-7}$&
          $6.3\times10^{-7}$&$6.3\times10^{-7}$&$6.3\times10^{-7}$\\
$^{36}$Ar&$3.9\times10^{-5}$&$3.9\times10^{-5}$&$3.9\times10^{-5}$&
          $3.9\times10^{-5}$&$3.9\times10^{-5}$&$3.9\times10^{-5}$&
          $3.9\times10^{-5}$&$3.9\times10^{-5}$&$3.9\times10^{-5}$&
          $3.9\times10^{-5}$&$3.9\times10^{-5}$&$3.9\times10^{-5}$\\
%$^{37}$Ar&$4.9\times10^{-15}$&$3.2\times10^{-14}$&$7.6\times10^{-14}$&
%          $1.3\times10^{-15}$&$1.7\times10^{-15}$&$1.9\times10^{-15}$&
%          $1.1\times10^{-17}$&$1.3\times10^{-15}$&$1.5\times10^{-17}$&
%          $9.5\times10^{-16}$&$1.3\times10^{-15}$&$1.5\times10^{-15}$\\
$^{38}$Ar&$5.6\times10^{-5}$&$5.6\times10^{-5}$&$5.6\times10^{-5}$&
          $5.6\times10^{-5}$&$5.2\times10^{-5}$&$5.6\times10^{-5}$&
          $5.5\times10^{-5}$&$5.4\times10^{-5}$&$5.4\times10^{-5}$&
          $5.5\times10^{-5}$&$5.4\times10^{-5}$&$5.2\times10^{-5}$\\
%$^{38}$K&$5.5\times10^{-16}$&$1.3\times10^{-17}$&---&
%          $3.1\times10^{-16}$&$1.9\times10^{-18}$&$2.1\times10^{-20}$&
%          $4.2\times10^{-16}$&$2.1\times10^{-18}$&$1.9\times10^{-20}$&
%          $3.0\times10^{-16}$&$2.1\times10^{-18}$&$2.2\times10^{-20}$\\
$^{39}$K&$3.7\times10^{-6}$&$3.7\times10^{-6}$&$3.9\times10^{-6}$&
         $3.7\times10^{-6}$&$3.5\times10^{-6}$&$3.7\times10^{-6}$&
         $3.6\times10^{-6}$&$3.6\times10^{-6}$&$3.6\times10^{-6}$&
         $3.7\times10^{-6}$&$3.6\times10^{-6}$&$3.5\times10^{-6}$\\
$^{40}$Ca&$6.3\times10^{-5}$&$6.3\times10^{-5}$&$6.3\times10^{-5}$&
          $6.3\times10^{-5}$&$5.8\times10^{-5}$&$6.3\times10^{-5}$&
          $6.2\times10^{-5}$&$6.1\times10^{-5}$&$6.2\times10^{-5}$&
          $6.2\times10^{-5}$&$6.1\times10^{-5}$&$5.9\times10^{-5}$\\
%$^{41}$Ca&$8.9\times10^{-15}$&$5.9\times10^{-14}$&$1.3\times10^{-13}$&
%          $1.7\times10^{-15}$&$2.5\times10^{-15}$&$2.6\times10^{-15}$&
%          $1.4\times10^{-15}$&$1.8\times10^{-15}$&$2.1\times10^{-15}$&
%          $1.3\times10^{-15}$&$1.8\times10^{-15}$&$2.1\times10^{-15}$\\
\hline
 \label{tab:1.0}
\end{tabular}
\end{table*}
\end{landscape}

\begin{landscape}
\begin{table*}
  \caption{Similar with Table \ref{tab:0.6} but for models with 1.2 $\rm M_\odot$ ONe WD£¬
  and $\Delta M_{\rm en}$ and $\Delta M_{\rm ej}$ are in unit of $10^{-6}\rm M_\odot$.}
  \tabcolsep0.1mm
  \begin{tabular}{ccccccccccccc}
\cline{1-13}
\multicolumn{1}{|c|}{}&\multicolumn{12}{|c|}{$M_{\rm WD}=1.2M_\odot$ } \\
\multicolumn{1}{|c|}{}&\multicolumn{3}{|c|}{$\delta=0.001$ }& \multicolumn{3}{|c|}{$\delta=0.01$ }&
\multicolumn{3}{|c|}{$\delta=0.05$ }&\multicolumn{3}{|c|}{$\delta=0.1$ }\\
\multicolumn{1}{|c|}{Models}&\multicolumn{1}{|c|}{$\dot{M}=10^{-7}$}&\multicolumn{1}{|c|}{$\dot{M}=10^{-9}$}
&\multicolumn{1}{|c|}{$\dot{M}=10^{-11}$}&\multicolumn{1}{|c|}{$\dot{M}=10^{-7}$}&\multicolumn{1}{|c|}{$\dot{M}=10^{-9}$}
&\multicolumn{1}{|c|}{$\dot{M}=10^{-11}$}&\multicolumn{1}{|c|}{$\dot{M}=10^{-7}$}&\multicolumn{1}{|c|}{$\dot{M}=10^{-9}$}
&\multicolumn{1}{|c|}{$\dot{M}=10^{-11}$}&\multicolumn{1}{|c|}{$\dot{M}=10^{-7}$}&\multicolumn{1}{|c|}{$\dot{M}=10^{-9}$}
&\multicolumn{1}{|c|}{$\dot{M}=10^{-11}$}\\

\cline{1-13} Results&1&2&3&4&5&6&7&8&9&10&11&12\\
$\Delta M_{\rm env}$&1.0&2.6&6.0&0.9&2.3&5.4&1.2&3.0&7.1&1.4&3.3&8.0\\
$\Delta M_{\rm eje}$&0.8&2.2&5.7&0.8&2.0&5.1&0.8&2.6&6.6&0.9&2.8&7.4\\
$T_{\rm max}$&1.3&1.6&2.0&1.2&1.5&1.9&1.4&1.7&2.0&1.4&1.7&2.1\\
$^{1}$H&$3.1\times10^{-1}$&$3.0\times10^{-1}$&$2.6\times10^{-1}$&
        $3.2\times10^{-1}$&$3.0\times10^{-1}$&$2.5\times10^{-1}$&
        $3.4\times10^{-1}$&$3.0\times10^{-1}$&$2.8\times10^{-1}$&
        $3.4\times10^{-1}$&$3.0\times10^{-1}$&$2.8\times10^{-1}$\\
$^{2}$H&$5.6\times10^{-11}$&$1.1\times10^{-11}$&$5.6\times10^{-14}$&
        $5.5\times10^{-11}$&$1.9\times10^{-12}$&$8.0\times10^{-14}$&
        $5.7\times10^{-11}$&$5.7\times10^{-12}$&$3.7\times10^{-13}$&
        $5.3\times10^{-11}$&$1.3\times10^{-12}$&$1.1\times10^{-12}$\\
$^{3}$He&$9.6\times10^{-7}$&$6.3\times10^{-7}$&$1.8\times10^{-7}$&
         $5.0\times10^{-6}$&$1.8\times10^{-6}$&$9.5\times10^{-7}$&
         $1.8\times10^{-6}$&$7.5\times10^{-8}$&$2.2\times10^{-8}$&
         $9.3\times10^{-7}$&$2.0\times10^{-8}$&$2.8\times10^{-9}$\\
$^{4}$He&$2.7\times10^{-1}$&$2.9\times10^{-1}$&$3.4\times10^{-1}$&
         $1.6\times10^{-1}$&$1.8\times10^{-1}$&$2.3\times10^{-1}$&
         $1.5\times10^{-1}$&$1.9\times10^{-1}$&$2.2\times10^{-1}$&
         $1.5\times10^{-1}$&$1.9\times10^{-1}$&$2.2\times10^{-1}$\\
%$^{6}$Li&$2.0\times10^{-14}$&$3.4\times10^{-16}$&$1.6\times10^{-18}$&
%         $2.0\times10^{-14}$&$5.3\times10^{-17}$&$2.2\times10^{-18}$&
%         $2.0\times10^{-14}$&$1.8\times10^{-16}$&$1.2\times10^{-17}$&
%         $1.8\times10^{-14}$&$3.7\times10^{-17}$&$3.9\times10^{-17}$\\
$^{7}$Li&$5.1\times10^{-11}$&$4.7\times10^{-11}$&$1.1\times10^{-11}$&
         $3.0\times10^{-11}$&$3.2\times10^{-11}$&$6.5\times10^{-12}$&
         $3.8\times10^{-11}$&$4.3\times10^{-11}$&$1.3\times10^{-11}$&
         $4.0\times10^{-11}$&$4.3\times10^{-11}$&$1.2\times10^{-11}$\\
$^{7}$Be&$2.0\times10^{-5}$&$2.0\times10^{-5}$&$1.0\times10^{-5}$&
         $1.2\times10^{-5}$&$1.4\times10^{-5}$&$8.5\times10^{-6}$&
         $1.3\times10^{-5}$&$1.3\times10^{-5}$&$8.5\times10^{-6}$&
         $1.4\times10^{-5}$&$1.2\times10^{-5}$&$7.9\times10^{-6}$\\
%$^{8}$Be&---&---&---&---&---&---&---&---&---&---&---&---\\
%$^{10}$B&$1.1\times10^{-13}$&$5.0\times10^{-16}$&$2.5\times10^{-18}$&
%         $7.8\times10^{-14}$&$7.8\times10^{-17}$&$3.3\times10^{-18}$&
%         $5.0\times10^{-14}$&$2.7\times10^{-16}$&$1.7\times10^{-17}$&
%         $3.8\times10^{-14}$&$5.4\times10^{-17}$&$5.8\times10^{-17}$\\
%$^{11}$B&$4.8\times10^{-13}$&$2.3\times10^{-15}$&$1.1\times10^{-17}$&
%         $3.5\times10^{-13}$&$3.5\times10^{-16}$&$1.5\times10^{-17}$&
%         $2.3\times10^{-13}$&$1.2\times10^{-15}$&$7.7\times10^{-17}$&
%         $1.7\times10^{-13}$&$2.4\times10^{-16}$&$2.6\times10^{-16}$\\
$^{12}$C&$8.8\times10^{-3}$&$9.8\times10^{-3}$&$2.5\times10^{-2}$&
         $1.9\times10^{-2}$&$1.2\times10^{-2}$&$4.2\times10^{-2}$&
         $3.2\times10^{-3}$&$4.4\times10^{-3}$&$7.5\times10^{-3}$&
         $2.1\times10^{-3}$&$3.3\times10^{-3}$&$6.6\times10^{-3}$\\
$^{13}$C&$2.7\times10^{-3}$&$3.1\times10^{-3}$&$1.5\times10^{-2}$&
         $6.6\times10^{-3}$&$3.8\times10^{-3}$&$3.2\times10^{-2}$&
         $9.6\times10^{-4}$&$1.3\times10^{-3}$&$2.9\times10^{-3}$&
         $6.2\times10^{-4}$&$1.0\times10^{-3}$&$2.5\times10^{-3}$\\
$^{14}$N&$2.1\times10^{-1}$&$2.0\times10^{-1}$&$1.8\times10^{-1}$&
         $2.3\times10^{-1}$&$2.5\times10^{-1}$&$1.9\times10^{-1}$&
         $8.1\times10^{-2}$&$1.1\times10^{-1}$&$1.2\times10^{-1}$&
         $5.4\times10^{-2}$&$8.1\times10^{-2}$&$1.1\times10^{-1}$\\
$^{15}$N&$1.2\times10^{-5}$&$1.1\times10^{-5}$&$2.8\times10^{-4}$&
         $7.6\times10^{-6}$&$1.5\times10^{-5}$&$3.7\times10^{-4}$&
         $5.0\times10^{-6}$&$1.3\times10^{-5}$&$5.5\times10^{-5}$&
         $4.8\times10^{-6}$&$1.2\times10^{-5}$&$5.5\times10^{-5}$\\
$^{16}$O&$1.8\times10^{-1}$&$1.8\times10^{-1}$&$1.7\times10^{-1}$&
         $2.4\times10^{-1}$&$2.3\times10^{-1}$&$2.2\times10^{-1}$&
         $2.0\times10^{-1}$&$1.8\times10^{-1}$&$1.5\times10^{-1}$&
         $2.0\times10^{-1}$&$1.7\times10^{-1}$&$1.2\times10^{-1}$\\
$^{17}$O&$1.5\times10^{-3}$&$2.6\times10^{-3}$&$5.9\times10^{-3}$&
         $1.7\times10^{-3}$&$3.2\times10^{-3}$&$7.8\times10^{-3}$&
         $1.7\times10^{-3}$&$2.0\times10^{-3}$&$3.8\times10^{-3}$&
         $1.7\times10^{-3}$&$1.8\times10^{-3}$&$3.1\times10^{-3}$\\
$^{18}$O&$1.7\times10^{-6}$&$4.0\times10^{-6}$&$9.7\times10^{-5}$&
         $5.7\times10^{-7}$&$4.3\times10^{-6}$&$6.9\times10^{-5}$&
         $1.8\times10^{-6}$&$1.3\times10^{-5}$&$8.6\times10^{-5}$&
         $3.5\times10^{-6}$&$1.6\times10^{-5}$&$8.5\times10^{-5}$\\
$^{18}$F&$2.1\times10^{-7}$&$1.3\times10^{-6}$&$2.9\times10^{-6}$&
         $8.0\times10^{-8}$&$1.4\times10^{-6}$&$2.0\times10^{-6}$&
         $2.3\times10^{-7}$&$2.7\times10^{-6}$&$2.6\times10^{-6}$&
         $3.6\times10^{-7}$&$3.0\times10^{-6}$&$2.2\times10^{-6}$\\
$^{19}$F&$5.1\times10^{-7}$&$2.8\times10^{-6}$&$9.0\times10^{-6}$&
         $4.2\times10^{-7}$&$2.6\times10^{-6}$&$1.2\times10^{-6}$&
         $5.6\times10^{-7}$&$5.1\times10^{-7}$&$3.9\times10^{-6}$&
         $4.6\times10^{-7}$&$4.2\times10^{-7}$&$2.8\times10^{-6}$\\
$^{20}$Ne&$2.3\times10^{-3}$&$3.2\times10^{-3}$&$3.6\times10^{-3}$&
          $1.5\times10^{-2}$&$1.5\times10^{-2}$&$1.6\times10^{-2}$&
          $1.9\times10^{-1}$&$1.9\times10^{-1}$&$1.9\times10^{-1}$&
          $2.2\times10^{-1}$&$2.2\times10^{-1}$&$2.2\times10^{-1}$\\
$^{21}$Ne&$4.3\times10^{-6}$&$4.5\times10^{-6}$&$4.5\times10^{-6}$&
          $4.3\times10^{-6}$&$4.2\times10^{-6}$&$4.2\times10^{-6}$&
          $4.3\times10^{-6}$&$4.2\times10^{-6}$&$4.3\times10^{-6}$&
          $4.3\times10^{-6}$&$4.2\times10^{-6}$&$4.3\times10^{-6}$\\
$^{22}$Ne&$3.1\times10^{-3}$&$2.9\times10^{-3}$&$2.4\times10^{-3}$&
          $6.4\times10^{-3}$&$4.3\times10^{-3}$&$4.0\times10^{-3}$&
          $4.4\times10^{-3}$&$1.0\times10^{-3}$&$7.3\times10^{-4}$&
          $3.1\times10^{-3}$&$4.9\times10^{-4}$&$2.5\times10^{-4}$\\
$^{23}$Na&$3.0\times10^{-3}$&$2.5\times10^{-3}$&$8.6\times10^{-4}$&
          $2.9\times10^{-3}$&$3.4\times10^{-3}$&$1.3\times10^{-3}$&
          $5.1\times10^{-3}$&$2.4\times10^{-3}$&$4.4\times10^{-4}$&
          $5.7\times10^{-3}$&$1.5\times10^{-3}$&$1.6\times10^{-4}$\\
$^{24}$Mg&$1.6\times10^{-3}$&$2.0\times10^{-3}$&$3.6\times10^{-3}$&
          $2.3\times10^{-3}$&$2.6\times10^{-3}$&$4.6\times10^{-3}$&
          $2.5\times10^{-2}$&$2.6\times10^{-2}$&$3.0\times10^{-2}$&
          $2.8\times10^{-2}$&$3.0\times10^{-2}$&$3.3\times10^{-2}$\\
$^{25}$Mg&$7.2\times10^{-5}$&$7.5\times10^{-5}$&$7.4\times10^{-5}$&
          $7.0\times10^{-5}$&$6.9\times10^{-5}$&$7.0\times10^{-5}$&
          $7.1\times10^{-5}$&$7.0\times10^{-5}$&$7.0\times10^{-5}$&
          $7.0\times10^{-5}$&$7.0\times10^{-5}$&$7.1\times10^{-5}$\\
$^{26}$Mg&$8.7\times10^{-5}$&$8.3\times10^{-5}$&$2.6\times10^{-5}$&
          $7.9\times10^{-5}$&$7.2\times10^{-5}$&$3.5\times10^{-5}$&
          $7.9\times10^{-5}$&$5.4\times10^{-5}$&$5.3\times10^{-7}$&
          $7.7\times10^{-5}$&$4.3\times10^{-5}$&$2.5\times10^{-8}$\\
$^{27}$Al&$5.9\times10^{-5}$&$4.3\times10^{-5}$&$1.9\times10^{-5}$&
          $6.1\times10^{-5}$&$4.3\times10^{-5}$&$2.5\times10^{-5}$&
          $5.8\times10^{-5}$&$2.1\times10^{-5}$&$5.0\times10^{-7}$&
          $5.6\times10^{-5}$&$1.4\times10^{-5}$&$2.7\times10^{-8}$\\
$^{28}$Si&$7.0\times10^{-4}$&$7.4\times10^{-4}$&$7.4\times10^{-4}$&
          $6.9\times10^{-4}$&$6.9\times10^{-4}$&$7.0\times10^{-4}$&
          $7.0\times10^{-4}$&$6.9\times10^{-4}$&$7.1\times10^{-4}$&
          $6.9\times10^{-4}$&$7.0\times10^{-4}$&$7.1\times10^{-4}$\\
$^{29}$Si&$3.7\times10^{-5}$&$3.9\times10^{-5}$&$3.8\times10^{-5}$&
          $3.6\times10^{-5}$&$3.6\times10^{-5}$&$3.6\times10^{-5}$&
          $3.6\times10^{-5}$&$3.6\times10^{-5}$&$3.6\times10^{-5}$&
          $3.6\times10^{-5}$&$3.6\times10^{-5}$&$3.7\times10^{-5}$\\
$^{30}$Si&$2.5\times10^{-5}$&$2.6\times10^{-5}$&$2.6\times10^{-5}$&
          $2.5\times10^{-5}$&$2.5\times10^{-5}$&$2.5\times10^{-5}$&
          $2.5\times10^{-5}$&$2.5\times10^{-5}$&$2.5\times10^{-5}$&
          $2.5\times10^{-5}$&$2.5\times10^{-5}$&$2.5\times10^{-5}$\\
$^{31}$P&$7.9\times10^{-6}$&$8.3\times10^{-6}$&$8.0\times10^{-6}$&
         $7.8\times10^{-6}$&$7.6\times10^{-6}$&$7.7\times10^{-6}$&
         $7.8\times10^{-6}$&$7.6\times10^{-6}$&$7.4\times10^{-6}$&
         $7.8\times10^{-6}$&$7.6\times10^{-6}$&$7.1\times10^{-6}$\\
$^{32}$S&$3.9\times10^{-4}$&$4.1\times10^{-4}$&$4.0\times10^{-4}$&
         $3.8\times10^{-4}$&$3.8\times10^{-4}$&$3.8\times10^{-4}$&
         $3.8\times10^{-4}$&$3.8\times10^{-4}$&$3.8\times10^{-4}$&
         $3.8\times10^{-4}$&$3.8\times10^{-4}$&$3.8\times10^{-4}$\\
$^{33}$S&$3.2\times10^{-6}$&$3.3\times10^{-6}$&$3.3\times10^{-6}$&
         $3.1\times10^{-6}$&$3.1\times10^{-6}$&$3.1\times10^{-6}$&
         $3.1\times10^{-6}$&$3.1\times10^{-6}$&$3.1\times10^{-6}$&
         $3.1\times10^{-6}$&$3.1\times10^{-6}$&$3.1\times10^{-6}$\\
$^{34}$S&$1.8\times10^{-5}$&$1.9\times10^{-5}$&$1.9\times10^{-5}$&
         $1.8\times10^{-5}$&$1.8\times10^{-5}$&$1.8\times10^{-5}$&
         $1.8\times10^{-5}$&$1.8\times10^{-5}$&$1.8\times10^{-5}$&
         $1.8\times10^{-5}$&$1.8\times10^{-5}$&$1.8\times10^{-5}$\\
$^{35}$Cl&$3.2\times10^{-6}$&$3.4\times10^{-6}$&$3.4\times10^{-6}$&
          $3.2\times10^{-6}$&$3.2\times10^{-6}$&$3.2\times10^{-6}$&
          $3.2\times10^{-6}$&$3.2\times10^{-6}$&$3.5\times10^{-6}$&
          $3.2\times10^{-6}$&$3.3\times10^{-6}$&$3.8\times10^{-6}$\\
%$^{36}$Cl&$6.2\times10^{-15}$&$1.0\times10^{-15}$&$1.8\times10^{-15}$&
%          $3.4\times10^{-15}$&$6.3\times10^{-16}$&$6.6\times10^{-16}$&
%          $2.6\times10^{-15}$&$9.6\times10^{-16}$&$1.5\times10^{-15}$&
%          $1.9\times10^{-15}$&$1.0\times10^{-15}$&$1.8\times10^{-15}$\\
$^{37}$Cl&$6.3\times10^{-7}$&$6.3\times10^{-7}$&$6.3\times10^{-7}$&
          $6.3\times10^{-7}$&$6.3\times10^{-7}$&$6.3\times10^{-7}$&
          $6.3\times10^{-7}$&$6.3\times10^{-7}$&$6.3\times10^{-7}$&
          $6.3\times10^{-7}$&$6.3\times10^{-7}$&$6.3\times10^{-7}$\\
$^{36}$Ar&$3.9\times10^{-5}$&$3.9\times10^{-5}$&$3.9\times10^{-5}$&
          $3.9\times10^{-5}$&$3.9\times10^{-5}$&$3.9\times10^{-5}$&
          $3.9\times10^{-5}$&$3.9\times10^{-5}$&$3.9\times10^{-5}$&
          $3.9\times10^{-5}$&$3.9\times10^{-5}$&$3.9\times10^{-5}$\\
%$^{37}$Ar&$7.6\times10^{-15}$&$9.1\times10^{-15}$&$1.5\times10^{-14}$&
%          $4.2\times10^{-15}$&$5.8\times10^{-15}$&$6.0\times10^{-15}$&
%          $3.5\times10^{-15}$&$8.8\times10^{-15}$&$1.3\times10^{-14}$&
%          $3.1\times10^{-15}$&$9.3\times10^{-15}$&$1.4\times10^{-14}$\\
$^{38}$Ar&$5.7\times10^{-5}$&$6.2\times10^{-5}$&$6.0\times10^{-5}$&
          $5.6\times10^{-5}$&$5.4\times10^{-5}$&$5.5\times10^{-5}$&
          $5.6\times10^{-5}$&$5.5\times10^{-5}$&$5.6\times10^{-5}$&
          $5.6\times10^{-5}$&$5.5\times10^{-5}$&$5.6\times10^{-5}$\\
%$^{38}$K&$1.5\times10^{-15}$&$2.2\times10^{-18}$&$2.3\times10^{-20}$&
%          $1.6\times10^{-15}$&$1.2\times10^{-18}$&$1.7\times10^{-20}$&
%          $1.2\times10^{-15}$&$1.6\times10^{-18}$&$3.3\times10^{-20}$&
%          $8.8\times10^{-16}$&$1.3\times10^{-18}$&$6.2\times10^{-20}$\\
$^{39}$K&$3.7\times10^{-6}$&$3.9\times10^{-6}$&$3.9\times10^{-6}$&
         $3.7\times10^{-6}$&$3.6\times10^{-6}$&$3.7\times10^{-6}$&
         $3.7\times10^{-6}$&$3.6\times10^{-6}$&$3.8\times10^{-6}$&
         $3.7\times10^{-6}$&$3.7\times10^{-6}$&$3.8\times10^{-6}$\\
$^{40}$Ca&$6.4\times10^{-5}$&$6.7\times10^{-5}$&$6.6\times10^{-5}$&
          $6.3\times10^{-5}$&$6.2\times10^{-5}$&$6.3\times10^{-5}$&
          $6.3\times10^{-5}$&$6.2\times10^{-5}$&$6.3\times10^{-5}$&
          $6.3\times10^{-5}$&$6.3\times10^{-5}$&$6.3\times10^{-5}$\\
%$^{41}$Ca&$1.5\times10^{-14}$&$1.2\times10^{-14}$&$2.0\times10^{-14}$&
%          $7.2\times10^{-15}$&$7.2\times10^{-15}$&$7.5\times10^{-15}$&
%          $5.7\times10^{-15}$&$1.1\times10^{-14}$&$1.6\times10^{-14}$&
%          $4.4\times10^{-15}$&$1.2\times10^{-14}$&$1.8\times10^{-14}$\\
\hline
 \label{tab:1.2}
\end{tabular}
\end{table*}
\end{landscape}

\begin{landscape}
\begin{table*}
  \caption{Similar with Table \ref{tab:0.6} but for models with 1.3 $\rm M_\odot$ ONe WD£¬
  and $\Delta M_{\rm en}$ and $\Delta M_{\rm ej}$ are in unit of $10^{-6}\rm M_\odot$. }
  \tabcolsep0.1mm
  \begin{tabular}{ccccccccccccc}
\cline{1-13}
\multicolumn{1}{|c|}{}&\multicolumn{12}{|c|}{$M_{\rm WD}=1.3M_\odot$ } \\
\multicolumn{1}{|c|}{}&\multicolumn{3}{|c|}{$\delta=0.001$ }& \multicolumn{3}{|c|}{$\delta=0.01$ }&
\multicolumn{3}{|c|}{$\delta=0.05$ }&\multicolumn{3}{|c|}{$\delta=0.1$ }\\
\multicolumn{1}{|c|}{Models}&\multicolumn{1}{|c|}{$\dot{M}=10^{-7}$}&\multicolumn{1}{|c|}{$\dot{M}=10^{-9}$}
&\multicolumn{1}{|c|}{$\dot{M}=10^{-11}$}&\multicolumn{1}{|c|}{$\dot{M}=10^{-7}$}&\multicolumn{1}{|c|}{$\dot{M}=10^{-9}$}
&\multicolumn{1}{|c|}{$\dot{M}=10^{-11}$}&\multicolumn{1}{|c|}{$\dot{M}=10^{-7}$}&\multicolumn{1}{|c|}{$\dot{M}=10^{-9}$}
&\multicolumn{1}{|c|}{$\dot{M}=10^{-11}$}&\multicolumn{1}{|c|}{$\dot{M}=10^{-7}$}&\multicolumn{1}{|c|}{$\dot{M}=10^{-9}$}
&\multicolumn{1}{|c|}{$\dot{M}=10^{-11}$}\\

\cline{1-13} Results&1&2&3&4&5&6&7&8&9&10&11&12\\
$\Delta M_{\rm env}$&0.9&1.2&3.1&1.2&1.7&3.8&1.2&2.0&4.0&1.2&2.1&4.1\\
$\Delta M_{\rm eje}$&0.7&1.1&2.9&0.9&1.5&3.8&1.0&1.9&3.9&1.0&1.9&3.9\\
$T_{\rm max}$&1.9&2.0&2.4&1.9&2.1&2.5&1.9&2.2&2.5&1.9&2.2&2.5\\
$^{1}$H&$2.8\times10^{-1}$&$2.5\times10^{-1}$&$2.3\times10^{-1}$&
        $2.7\times10^{-1}$&$2.6\times10^{-1}$&$2.3\times10^{-1}$&
        $2.7\times10^{-1}$&$2.4\times10^{-1}$&$2.3\times10^{-1}$&
        $2.7\times10^{-1}$&$2.5\times10^{-1}$&$2.3\times10^{-1}$\\
$^{2}$H&$2.5\times10^{-12}$&$1.1\times10^{-12}$&$6.2\times10^{-13}$&
        $4.9\times10^{-12}$&$9.8\times10^{-13}$&$6.5\times10^{-12}$&
        $5.0\times10^{-12}$&$1.0\times10^{-12}$&$1.2\times10^{-12}$&
        $2.7\times10^{-12}$&$8.9\times10^{-13}$&$8.4\times10^{-13}$\\
$^{3}$He&$2.6\times10^{-9}$&$4.9\times10^{-10}$&$1.7\times10^{-11}$&
         $2.2\times10^{-10}$&$4.2\times10^{-11}$&$3.7\times10^{-10}$&
         $1.7\times10^{-10}$&$1.8\times10^{-11}$&$2.3\times10^{-11}$&
         $1.5\times10^{-10}$&$1.6\times10^{-11}$&$2.4\times10^{-11}$\\
$^{4}$He&$2.2\times10^{-1}$&$2.5\times10^{-1}$&$2.8\times10^{-1}$&
         $2.3\times10^{-1}$&$2.5\times10^{-1}$&$2.9\times10^{-1}$&
         $2.4\times10^{-1}$&$2.7\times10^{-1}$&$2.9\times10^{-1}$&
         $2.4\times10^{-1}$&$2.6\times10^{-1}$&$2.9\times10^{-1}$\\
%$^{6}$Li&$6.8\times10^{-17}$&$3.2\times10^{-17}$&$2.3\times10^{-17}$&
%         $1.5\times10^{-16}$&$2.8\times10^{-17}$&$4.4\times10^{-16}$&
%         $1.5\times10^{-16}$&$5.3\times10^{-17}$&$7.7\times10^{-17}$&
%         $7.4\times10^{-17}$&$3.0\times10^{-17}$&$6.4\times10^{-17}$\\
$^{7}$Li&$2.0\times10^{-11}$&$1.3\times10^{-11}$&$2.5\times10^{-12}$&
         $1.7\times10^{-11}$&$9.3\times10^{-12}$&$1.8\times10^{-12}$&
         $1.5\times10^{-11}$&$6.2\times10^{-12}$&$1.4\times10^{-12}$&
         $1.5\times10^{-11}$&$6.1\times10^{-12}$&$1.4\times10^{-12}$\\
$^{7}$Be&$1.5\times10^{-5}$&$1.3\times10^{-5}$&$7.7\times10^{-6}$&
         $1.4\times10^{-5}$&$1.0\times10^{-5}$&$6.4\times10^{-6}$&
         $1.3\times10^{-5}$&$9.2\times10^{-6}$&$5.5\times10^{-6}$&
         $1.3\times10^{-5}$&$8.8\times10^{-6}$&$5.9\times10^{-6}$\\
%$^{8}$Be&---&---&---&---&---&---&---&---&---&---&---&---\\
%$^{10}$B&$1.0\times10^{-16}$&$4.8\times10^{-17}$&$3.4\times10^{-17}$&
%         $2.2\times10^{-16}$&$4.1\times10^{-17}$&$9.0\times10^{-16}$&
%         $2.3\times10^{-16}$&$7.9\times10^{-17}$&$1.2\times10^{-16}$&
%         $1.1\times10^{-16}$&$4.5\times10^{-17}$&$1.0\times10^{-16}$\\
%$^{11}$B&$4.5\times10^{-16}$&$2.1\times10^{-16}$&$1.5\times10^{-16}$&
%         $9.7\times10^{-16}$&$1.8\times10^{-16}$&$4.0\times10^{-15}$&
 %        $1.0\times10^{-15}$&$3.5\times10^{-16}$&$5.2\times10^{-16}$&
 %        $4.9\times10^{-16}$&$2.0\times10^{-16}$&$4.5\times10^{-16}$\\
$^{12}$C&$3.9\times10^{-3}$&$5.2\times10^{-3}$&$2.0\times10^{-2}$&
         $4.0\times10^{-3}$&$5.6\times10^{-3}$&$2.8\times10^{-2}$&
         $4.1\times10^{-3}$&$8.4\times10^{-3}$&$3.2\times10^{-2}$&
         $4.1\times10^{-3}$&$7.6\times10^{-3}$&$3.3\times10^{-2}$\\
$^{13}$C&$1.2\times10^{-3}$&$1.7\times10^{-3}$&$1.6\times10^{-2}$&
         $1.3\times10^{-3}$&$1.9\times10^{-3}$&$2.6\times10^{-2}$&
         $1.3\times10^{-3}$&$3.2\times10^{-3}$&$3.1\times10^{-2}$&
         $1.3\times10^{-3}$&$2.8\times10^{-3}$&$3.2\times10^{-2}$\\
$^{14}$N&$8.2\times10^{-2}$&$1.0\times10^{-1}$&$1.3\times10^{-1}$&
         $8.2\times10^{-2}$&$1.0\times10^{-1}$&$1.2\times10^{-1}$&
         $8.4\times10^{-2}$&$1.3\times10^{-1}$&$1.1\times10^{-1}$&
         $8.4\times10^{-2}$&$1.2\times10^{-1}$&$1.1\times10^{-1}$\\
$^{15}$N&$2.6\times10^{-5}$&$3.0\times10^{-5}$&$6.3\times10^{-4}$&
         $2.7\times10^{-5}$&$3.8\times10^{-5}$&$3.6\times10^{-3}$&
         $2.7\times10^{-5}$&$6.4\times10^{-5}$&$4.3\times10^{-3}$&
         $2.7\times10^{-5}$&$5.7\times10^{-5}$&$4.7\times10^{-3}$\\
$^{16}$O&$1.2\times10^{-1}$&$1.0\times10^{-1}$&$2.5\times10^{-2}$&
         $1.2\times10^{-1}$&$9.1\times10^{-2}$&$9.9\times10^{-3}$&
         $1.2\times10^{-1}$&$6.8\times10^{-2}$&$8.3\times10^{-3}$&
         $1.2\times10^{-1}$&$7.1\times10^{-2}$&$7.4\times10^{-3}$\\
$^{17}$O&$1.8\times10^{-3}$&$1.8\times10^{-3}$&$2.2\times10^{-3}$&
         $2.0\times10^{-3}$&$1.8\times10^{-3}$&$2.4\times10^{-3}$&
         $1.9\times10^{-3}$&$1.8\times10^{-3}$&$2.6\times10^{-3}$&
         $1.9\times10^{-3}$&$1.7\times10^{-3}$&$2.6\times10^{-3}$\\
$^{18}$O&$2.9\times10^{-5}$&$2.0\times10^{-5}$&$1.7\times10^{-4}$&
         $2.3\times10^{-5}$&$2.9\times10^{-5}$&$1.1\times10^{-4}$&
         $2.2\times10^{-5}$&$5.1\times10^{-5}$&$1.2\times10^{-4}$&
         $2.2\times10^{-5}$&$4.7\times10^{-5}$&$1.1\times10^{-4}$\\
$^{18}$F&$1.3\times10^{-5}$&$1.1\times10^{-5}$&$1.6\times10^{-4}$&
         $1.2\times10^{-5}$&$8.5\times10^{-6}$&$2.4\times10^{-4}$&
         $1.2\times10^{-5}$&$7.4\times10^{-6}$&$3.2\times10^{-4}$&
         $1.2\times10^{-5}$&$7.2\times10^{-6}$&$3.2\times10^{-4}$\\
$^{19}$F&$3.8\times10^{-7}$&$4.4\times10^{-7}$&$1.9\times10^{-6}$&
         $4.1\times10^{-7}$&$4.9\times10^{-7}$&$1.3\times10^{-6}$&
         $4.1\times10^{-7}$&$9.2\times10^{-7}$&$1.2\times10^{-6}$&
         $4.1\times10^{-7}$&$7.7\times10^{-7}$&$1.1\times10^{-6}$\\
$^{20}$Ne&$2.5\times10^{-1}$&$2.4\times10^{-1}$&$2.3\times10^{-1}$&
          $2.5\times10^{-1}$&$2.4\times10^{-1}$&$2.1\times10^{-1}$&
          $2.5\times10^{-1}$&$2.3\times10^{-1}$&$2.1\times10^{-1}$&
          $2.5\times10^{-1}$&$2.4\times10^{-1}$&$2.1\times10^{-1}$\\
$^{21}$Ne&$4.3\times10^{-6}$&$4.2\times10^{-6}$&$4.3\times10^{-6}$&
          $4.3\times10^{-6}$&$4.3\times10^{-6}$&$4.3\times10^{-6}$&
          $4.3\times10^{-6}$&$4.2\times10^{-6}$&$4.3\times10^{-6}$&
          $4.3\times10^{-6}$&$4.3\times10^{-6}$&$4.3\times10^{-6}$\\
$^{22}$Ne&$2.4\times10^{-4}$&$1.8\times10^{-4}$&$4.1\times10^{-6}$&
          $6.8\times10^{-5}$&$3.6\times10^{-5}$&$1.9\times10^{-7}$&
          $5.6\times10^{-5}$&$1.1\times10^{-5}$&$3.2\times10^{-8}$&
          $5.7\times10^{-5}$&$1.4\times10^{-5}$&$2.4\times10^{-8}$\\
$^{23}$Na&$4.8\times10^{-4}$&$2.0\times10^{-4}$&$1.0\times10^{-6}$&
          $1.1\times10^{-4}$&$3.2\times10^{-5}$&$4.3\times10^{-8}$&
          $8.5\times10^{-5}$&$5.8\times10^{-6}$&$5.4\times10^{-9}$&
          $8.3\times10^{-5}$&$8.7\times10^{-6}$&$4.0\times10^{-9}$\\
$^{24}$Mg&$3.6\times10^{-2}$&$3.6\times10^{-2}$&$3.8\times10^{-2}$&
          $3.4\times10^{-2}$&$3.4\times10^{-2}$&$3.4\times10^{-2}$&
          $3.4\times10^{-2}$&$3.3\times10^{-2}$&$3.3\times10^{-2}$&
          $3.4\times10^{-2}$&$3.3\times10^{-2}$&$3.3\times10^{-2}$\\
$^{25}$Mg&$7.1\times10^{-5}$&$6.9\times10^{-5}$&$7.1\times10^{-5}$&
          $7.1\times10^{-5}$&$7.1\times10^{-5}$&$7.1\times10^{-5}$&
          $7.1\times10^{-5}$&$7.0\times10^{-5}$&$7.1\times10^{-5}$&
          $7.1\times10^{-5}$&$7.0\times10^{-5}$&$7.1\times10^{-5}$\\
$^{26}$Mg&$1.1\times10^{-5}$&$2.1\times10^{-7}$&$7.0\times10^{-11}$&
          $4.2\times10^{-7}$&$3.6\times10^{-10}$&$1.3\times10^{-9}$&
          $1.8\times10^{-7}$&$9.0\times10^{-11}$&$3.9\times10^{-11}$&
          $1.5\times10^{-7}$&$1.2\times10^{-10}$&$3.9\times10^{-11}$\\
$^{27}$Al&$3.6\times10^{-6}$&$1.2\times10^{-7}$&$5.5\times10^{-11}$&
          $1.7\times10^{-7}$&$2.4\times10^{-10}$&$1.0\times10^{-9}$&
          $7.7\times10^{-8}$&$5.4\times10^{-11}$&$2.8\times10^{-11}$&
          $6.8\times10^{-8}$&$7.1\times10^{-11}$&$2.9\times10^{-11}$\\
$^{28}$Si&$7.1\times10^{-4}$&$7.0\times10^{-4}$&$7.1\times10^{-4}$&
          $7.1\times10^{-4}$&$7.1\times10^{-4}$&$7.1\times10^{-4}$&
          $7.1\times10^{-4}$&$7.0\times10^{-4}$&$7.1\times10^{-4}$&
          $7.1\times10^{-4}$&$7.1\times10^{-4}$&$7.1\times10^{-4}$\\
$^{29}$Si&$3.7\times10^{-5}$&$3.6\times10^{-5}$&$3.7\times10^{-5}$&
          $3.7\times10^{-5}$&$3.7\times10^{-5}$&$3.7\times10^{-5}$&
          $3.7\times10^{-5}$&$3.6\times10^{-5}$&$3.7\times10^{-5}$&
          $3.7\times10^{-5}$&$3.6\times10^{-5}$&$3.7\times10^{-5}$\\
$^{30}$Si&$2.5\times10^{-5}$&$2.5\times10^{-5}$&$2.3\times10^{-5}$&
          $2.5\times10^{-5}$&$2.5\times10^{-5}$&$2.0\times10^{-5}$&
          $2.5\times10^{-5}$&$2.4\times10^{-5}$&$1.9\times10^{-5}$&
          $2.5\times10^{-5}$&$2.5\times10^{-5}$&$1.8\times10^{-5}$\\
$^{31}$P&$7.6\times10^{-6}$&$7.1\times10^{-6}$&$4.2\times10^{-6}$&
         $7.3\times10^{-6}$&$6.7\times10^{-6}$&$3.2\times10^{-6}$&
         $7.2\times10^{-6}$&$5.9\times10^{-6}$&$3.0\times10^{-6}$&
         $7.2\times10^{-6}$&$6.0\times10^{-6}$&$3.0\times10^{-6}$\\
$^{32}$S&$3.8\times10^{-4}$&$3.7\times10^{-4}$&$3.9\times10^{-4}$&
         $3.8\times10^{-4}$&$3.8\times10^{-4}$&$3.9\times10^{-4}$&
         $3.8\times10^{-4}$&$3.8\times10^{-4}$&$3.9\times10^{-4}$&
         $3.8\times10^{-4}$&$3.8\times10^{-4}$&$3.9\times10^{-4}$\\
$^{33}$S&$3.1\times10^{-6}$&$3.0\times10^{-6}$&$3.1\times10^{-6}$&
         $3.1\times10^{-6}$&$3.1\times10^{-6}$&$3.1\times10^{-6}$&
         $3.1\times10^{-6}$&$3.1\times10^{-6}$&$3.1\times10^{-6}$&
         $3.1\times10^{-6}$&$3.1\times10^{-6}$&$3.1\times10^{-6}$\\
$^{34}$S&$1.8\times10^{-5}$&$1.7\times10^{-5}$&$1.4\times10^{-5}$&
         $1.8\times10^{-5}$&$1.7\times10^{-5}$&$1.0\times10^{-5}$&
         $1.8\times10^{-5}$&$1.6\times10^{-5}$&$9.5\times10^{-6}$&
         $1.8\times10^{-5}$&$1.6\times10^{-5}$&$9.1\times10^{-6}$\\
$^{35}$Cl&$3.4\times10^{-6}$&$3.5\times10^{-6}$&$7.5\times10^{-6}$&
          $3.6\times10^{-6}$&$4.1\times10^{-6}$&$1.0\times10^{-5}$&
          $3.7\times10^{-6}$&$4.8\times10^{-6}$&$1.1\times10^{-5}$&
          $3.7\times10^{-6}$&$4.8\times10^{-6}$&$1.1\times10^{-5}$\\
%$^{36}$Cl&$3.2\times10^{-15}$&$5.3\times10^{-15}$&$1.1\times10^{-14}$&
%          $4.6\times10^{-15}$&$6.9\times10^{-15}$&$1.4\times10^{-14}$&
%          $4.9\times10^{-15}$&$8.8\times10^{-15}$&$1.3\times10^{-14}$&
 %         $5.0\times10^{-15}$&$8.1\times10^{-15}$&$1.3\times10^{-15}$\\
$^{37}$Cl&$6.3\times10^{-7}$&$6.3\times10^{-7}$&$5.9\times10^{-7}$&
          $6.3\times10^{-7}$&$6.3\times10^{-7}$&$5.4\times10^{-7}$&
          $6.3\times10^{-7}$&$6.2\times10^{-7}$&$5.3\times10^{-7}$&
          $6.3\times10^{-7}$&$6.2\times10^{-7}$&$5.2\times10^{-7}$\\
$^{36}$Ar&$3.9\times10^{-5}$&$3.9\times10^{-5}$&$3.9\times10^{-5}$&
          $3.9\times10^{-5}$&$3.9\times10^{-5}$&$4.0\times10^{-5}$&
          $3.9\times10^{-5}$&$3.9\times10^{-5}$&$4.0\times10^{-5}$&
          $3.9\times10^{-5}$&$3.9\times10^{-5}$&$4.1\times10^{-5}$\\
%$^{37}$Ar&$3.0\times10^{-14}$&$4.7\times10^{-14}$&$4.8\times10^{-14}$&
%          $4.1\times10^{-14}$&$5.2\times10^{-14}$&$4.2\times10^{-14}$&
%          $4.3\times10^{-14}$&$5.7\times10^{-14}$&$4.1\times10^{-14}$&
%          $4.3\times10^{-14}$&$5.3\times10^{-14}$&$4.0\times10^{-14}$\\
$^{38}$Ar&$5.6\times10^{-5}$&$5.4\times10^{-5}$&$5.5\times10^{-5}$&
          $5.6\times10^{-5}$&$5.6\times10^{-5}$&$5.3\times10^{-5}$&
          $5.6\times10^{-5}$&$5.5\times10^{-5}$&$5.3\times10^{-5}$&
          $5.6\times10^{-5}$&$5.5\times10^{-5}$&$5.3\times10^{-5}$\\
%$^{38}$K&$2.5\times10^{-15}$&$8.4\times10^{-18}$&$5.1\times10^{-18}$&
%          $1.6\times10^{-15}$&$1.8\times10^{-18}$&$8.7\times10^{-17}$&
%          $1.6\times10^{-15}$&$1.9\times10^{-18}$&$9.4\times10^{-18}$&
%          $1.6\times10^{-15}$&$1.6\times10^{-17}$&$4.1\times10^{-18}$\\
$^{39}$K&$3.7\times10^{-6}$&$3.7\times10^{-6}$&$5.1\times10^{-6}$&
         $3.8\times10^{-6}$&$3.9\times10^{-6}$&$6.7\times10^{-6}$&
         $3.8\times10^{-6}$&$4.1\times10^{-6}$&$7.2\times10^{-6}$&
         $3.8\times10^{-6}$&$4.1\times10^{-6}$&$7.4\times10^{-6}$\\
$^{40}$Ca&$6.3\times10^{-5}$&$6.1\times10^{-5}$&$6.3\times10^{-5}$&
          $6.3\times10^{-5}$&$6.3\times10^{-5}$&$6.4\times10^{-5}$&
          $6.3\times10^{-5}$&$6.2\times10^{-5}$&$6.4\times10^{-5}$&
          $6.3\times10^{-5}$&$6.3\times10^{-5}$&$6.4\times10^{-5}$\\
%$^{41}$Ca&$3.7\times10^{-14}$&$5.6\times10^{-14}$&$5.8\times10^{-14}$&
%          $5.0\times10^{-14}$&$6.5\times10^{-14}$&$5.1\times10^{-14}$&
%          $5.3\times10^{-14}$&$6.9\times10^{-14}$&$4.9\times10^{-14}$&
%          $5.4\times10^{-14}$&$6.6\times10^{-14}$&$4.8\times10^{-14}$\\
\hline
 \label{tab:cases}
\end{tabular}
\end{table*}
\end{landscape}
%\newpage %Just because of unusual number of tables stacked at end

\bibliographystyle{mn2e}
\bibliography{lglmn,lglapj}
%\begin{thebibliography}{99}
%\include{lgl.bib}
%\end{thebibliography}
\bsp
\label{lastpage}

\end{document}